\documentclass{article}

\usepackage{latexsym,amssymb}
\usepackage{graphics}
\usepackage{pstricks,psfrag}
\usepackage{amsmath}
\usepackage{subfigure}
\usepackage{vmargin}  
\usepackage{epsfig}

\newrgbcolor{myred}{1 0 0}
\newrgbcolor{mygreen}{0 1 0 }
\newrgbcolor{myblue}{0 0 1}

\setmarginsrb{2.cm}{2.cm}{2.cm}{2.5cm}{0cm}{0.cm}{0cm}{1.cm}
\begin{document}

\large

\title{\bf Neutrino Mass Patterns within the See-saw Model
from Multi-localization along Extra Dimensions} 
\author{J.-M. Fr\`ere \footnote{e-mail: {\tt frere@ulb.ac.be}} , 
\ G. Moreau \footnote{e-mail: {\tt Gregory.Moreau@ulb.ac.be}}
\ and \ E. Nezri \footnote{e-mail: {\tt nezri@in2p3.fr}} \\ \\
{\it Service de Physique Th\'eorique, CP 225}\\
{\it Universit\'e Libre de Bruxelles, B-1050, Brussels, Belgium}}
\maketitle

\begin{abstract} 
We study a multi-localization model for charged leptons 
and neutrinos, including the possibility of a see-saw mechanism. This 
framework offers the opportunity to allow for realistic solutions 
in a consistent model without fine-tuning of parameters, even if quarks 
are also considered. Those solutions predict that the large Majorana mass 
eigenvalues for right-handed neutrinos are of the same order of magnitude, 
although this almost common mass can span a large range (bounded from 
above by $\sim 10^{12}{\rm GeV}$).  
The model also predicts Majorana masses between $\sim 10^{-2}{\rm eV}$ and 
$\sim 5 \ 10^{-2}{\rm eV}$ for the left-handed neutrinos, both in the normal 
and inverted mass hierarchy cases. This mass interval corresponds to 
sensitivities which are reachable by proposed neutrinoless double 
$\beta$ decay experiments. The preferred range for leptonic mixing angle 
$\theta_{13}$ is: $10^{-2} \lesssim \sin \theta_{13} \lesssim 10^{-1}$,
but smaller values are not totally excluded by the model.
\end{abstract}

\vskip .5cm

PACS numbers: 11.10.Kk, 12.15.Ff, 14.60.Pq, 14.60.St

\section{Introduction}
\label{intro}

One of the profound mysteries in particle physics is the origin
of strong mass hierarchy existing among the different generations
of quarks and leptons.
The Standard Model (SM) generates the measured quark and lepton masses
with the dimensionless Yukawa couplings, which are spread over
numerous orders of magnitude, so that it does not provide
a real interpretation to the observed hierarchical pattern.  
Hence, the SM fermion mass hierarchy should be explained by an higher 
energy theory. The most famous example for such a theory is
certainly based on the Froggatt-Nielsen mechanism \cite{FNmech}.
This mechanism introduces a `flavor symmetry' forbidding most of 
the Yukawa interactions. However, effective Yukawa couplings are
generated by the spontaneous breaking of the additional symmetry.
Hence, those couplings are suppressed by different powers of the 
breaking scale over some fundamental high energy scale.

The recent renewed interest for the physics of extra dimensions
\cite{ADD,RS} has lead to  
approaches toward the SM fermion mass hierarchy problem
completely different from the conventional ones. Those new  
approaches are attractive as they do not rely on the 
existence of any new symmetry in the short-distance theory.  
For example,
in a framework inspired by the Randall-Sundrum model \cite{RS},
the large SM fermion mass hierarchy can be understood by means of the 
metric ``warp'' factors \cite{Gher,Hube,Dool}. 
The fermion mass hierarchy can also be generated naturally
by permitting the fermion masses to evolve with a power-law 
dependence on the mass scale, in theories with extra space-time
dimensions \cite{Dien1,Dien2}.
Other interesting ideas for solving the mass hierarchy 
problem with extra dimension(s) can be found in the literature 
\cite{Yosh,Band,Nero,FlavSymED}.

N.~Arkani-Hamed and M.~Schmaltz have suggested a particularly 
original and  
natural interpretation of the SM fermion mass hierarchy \cite{AS}. 
In the Arkani-Hamed-Schmaltz (AS) scenario, the SM fermions are 
localized at different positions along extra spatial dimension(s) 
in which our four-dimensional domain wall has thus a spread. 
This localization can be achieved by using 
either non-perturbative effects in string/M theory or field-theory 
methods. One possible field-theoretical mechanism is to couple the
SM fermions to scalar fields which have vacuum expectation
values depending on the extra dimension(s). Indeed, it is known that
chiral fermions are confined in solitonic backgrounds \cite{Jack}. 
One of the effects of having the SM fermions ``stuck'' at 
different points in the wall is the following: the relative 
displacements of the SM fermion wave function peaks produce
suppression factors in the effective four-dimensional Yukawa couplings. 
These suppression factors being determined by the overlaps of SM fermion 
wave functions (getting smaller as the distance between wave function
peaks increases), they can vary with the fermion flavors and thus 
generate the wanted mass hierarchy.

Some interesting variations of the AS scenario, in which the 
four-dimensional fermions appear as zero modes trapped in the 
core of a topological defect, have been studied in \cite{vortex}. 
The possibilities of localizing the Higgs field in extra dimension(s)
\cite{Dval} or having different fermion wave function widths \cite{Hung1},
that would modify the suppression factors arising in AS models,
have also been explored. Furthermore, the AS idea has been considered 
in the contexts of supersymmetry \cite{SUSY1,SUSY2} and the Grand 
Unified Theories (GUT) \cite{GUT1,GUT2}. Let us mention finally that the 
effects of gauge interactions on AS models could help in understanding 
the quark mass hierarchy \cite{Nuss}.

Concrete realizations of the AS scenario have been constructed
\cite{Mira,Bare,Bran,Hung2}, as we will discuss now. 
In the case of the existence of only one extra dimension,
it has been demonstrated that the experimental values for 
quark masses and $V_{CKM}$ matrix elements lead to a unique 
characteristic configuration of the field localization \cite{Mira}. 
Concerning the charged lepton sector, only one simple example of
wave function positions reproducing the measured charged lepton masses 
has been given in the literature \cite{Mira}. Let us now consider the 
situation where right-handed neutrinos are added to the SM so that 
neutrinos acquire ordinary Dirac masses after electroweak symmetry 
breaking, exactly like the quarks and charged leptons. Then, if these 
right-handed neutrinos are also localized in the domain wall, it is 
possible to find several field configurations yielding to appropriate 
neutrino masses and mixing angles for explaining all the
experimental neutrino data \cite{Bare}. However, in this case of 
Dirac masses for neutrinos, it turns out that the positions of
neutrino fields are closely related \cite{Bare}. This fine-tuning
problem for the fundamental parameters is mainly due to the large leptonic 
mixing angles required by neutrino oscillation solutions to the neutrino 
puzzle \cite{data}, and thus does not appear in the quark sector.

In the present work, we investigate the alternative possibility
that, within the AS scenario, neutrinos acquire masses of Majorana 
type (instead of Dirac type as previously studied). We concentrate
on the `see-saw' model \cite{seesaw} which constitutes probably the
most elegant way of generating Majorana neutrino masses. The whole 
leptonic sector is considered, since the neutrino and charged lepton
sectors are related phenomenologically through the data on leptonic 
mixing angles, and, theoretically via the field positions in the wall 
of left-handed $SU(2)_L$ doublets of leptons. Furthermore, we restrict 
ourselves to the minimal case where the domain wall is thick in only 
one of the extra dimensions.
\\ In this study, we address the question of the existence of 
field localization configurations fitting all the present experimental 
data on charged leptons and neutrinos. In other words, we are interested
in the structure of lepton flavor space. We will show that it is indeed 
possible to find wave function displacement configurations in agreement 
with experimental results, and we will present complete realistic AS models. 
\\ First, a general description of the less fine-tuned solutions reproducing
the charged lepton masses will be given. Then, within this context,
we will show that the fine-tuning of neutrino field positions, which has 
appeared in the case of Dirac masses for neutrinos \cite{Bare}, is softer 
in the see-saw framework considered here. The reasons will be exposed in 
details. This result that the fine-tuning reaches an acceptable level in
the whole leptonic sector, which was the most serious phenomenological 
challenge for the AS scenario \cite{Bare}, makes this kind of AS model 
a realistic candidate for the solution to the mass hierarchy problem.   
\\ Moreover, it will be pointed out that the studied model, namely
the see-saw mechanism within the AS scenario, gives rise to clear
predictions on the value of light neutrino masses and to favored
values for the leptonic mixing angle $\theta_{13}$. Those predictions 
are interesting since they will be testable in the next generation of 
terrestrial neutrino experiments, as will be discussed.
\\ Finally, we will illustrate that, within this model, the reduction 
of neutrino masses compared to the electroweak symmetry breaking scale 
can be due partially to the see-saw mechanism, and partially to the 
suppression factors issued from field localization effect. Therefore, 
the field positions can be closer to each other than in the case of 
neutrino Dirac masses where the mass reduction comes entirely from the 
localization effect. This is attractive for two reasons. 
First, it renders the upper bound on the 
wall width, coming from perturbativity considerations, easier to 
respect. Secondly, it pleads for the `naturality' of 
the AS scenario, in the sense that generating large mass 
hierarchies from field positions (namely 
combinations of the fundamental parameters) of the same order of 
magnitude can be considered as a satisfactory and natural property.
This property is maybe the main progress brought by the AS scenario 
with regard to the SM fermion mass hierarchy problem.

At this level, one should mention a preliminary study on the same 
subject, performed in \cite{Klap}, even if the approach adopted
was much less generic than here: the authors of \cite{Klap} 
have considered, within the AS scenario, the combination of the
see-saw mechanism together with a model containing a triplet Higgs 
scalar, in view of predicting a degenerate diagonal neutrino mass 
spectrum. Such a spectrum allows a neutrino
contribution to the hot component of the dark matter of the universe.
Let us note that the authors of \cite{Klap} have given the example 
of a model reproducing the experimental data on leptons, in the context
of the see-saw mechanism within the AS framework. However, 
this example concerns a two extra dimension version of the model, 
and is associated to the small
mixing angle solution of the solar neutrino problem which has been
ruled out by the recent experimental results \cite{data}.

In another interesting previous related work \cite{Raidal:2002xf}, the AS scenario 
has been mentioned as a possible natural framework, for justifying the 
neutrino texture of a particular well motivated
see-saw model. This approach of the AS scenario was thus purely effective, 
in the two following senses. First, the fundamental parameters, namely the 
field positions, were replaced by effective neutrino mass parameters. 
Secondly, the charged lepton sector, which is not independent of the neutrino 
sector within the AS framework (see above), was not fully treated. 
We will come back on the study of \cite{Raidal:2002xf} later.

A last comment may be done at this stage. As already said, the scenario we
will study can explain both the structure in neutrino flavor space and the
(partial) suppression of neutrino mass scales compared to the electroweak 
scale, in terms of geometrical patterns in an higher-dimensional space. There
exist two other types of scenarios based on the existence of extra 
dimension(s) which offer the opportunity to explain the smallness of neutrino 
masses. Nevertheless, these scenarios do not provide interpretations 
of the neutrino flavor structure. In the first type of scenario, the lepton number 
breaking occurs on distant branes and is conveyed to our `3-brane' by a scalar 
field, leading then to weak effective neutrino Majorana masses 
\cite{LightNu1,LightNu2,LightNu0}. In the other kind of model, the 
right-handed neutrinos live in the bulk which gives rise to small neutrino 
Dirac masses, for the same reason that gravity is weak at low energy in this 
context \cite{LightNu1,LightNu2}.

In Section \ref{sec:formalism}, we discuss 
the see-saw mechanism within the context of the AS framework. Then, in 
Sections \ref{sec:2flavors} and \ref{sec:3flavors}, we construct consistent 
realizations of the AS scenario reproducing all the present experimental data
on the whole leptonic sector, in the cases of 2 and 3 flavors respectively.
Finally, in Section \ref{subsec:predict}, we present predictions on neutrino
sector provided by the AS scenario based on the see-saw mechanism.

\section{The See-saw Mechanism within the AS Framework}
\label{sec:formalism}

\subsection{The AS scenario}
\label{subsec:AS}

We briefly recall here the physical context and formalism of 
the AS scenario.
The SM degrees of freedom live on a four-dimensional wall
embedded in an higher-dimensional space where gravity, and 
possibly other gauge singlet fields, are free to propagate.
We consider the simple case where the domain wall is slightly 
thick in only one extra dimension. Inside our wall, the Higgs 
and gauge bosons are free to propagate whereas the SM fermions
of each family 
are trapped at different points. In an effective field theory 
approach, this field localization can be due to the coupling of 
each fermion $\Psi_i(x_\mu,x_5) \ [i=1,...,3;\mu=1,...,4]$ to a 
five-dimensional scalar field having 
a vacuum expectation value $\Phi_i(x_5)$ which varies along 
the extra dimension (parameterized by $x_5$). Under the 
hypothesis that the scalar field profile behaves as a linear function 
of the type $\Phi_i(x_5)=2\mu^2x_5-m_i$ around its zero-crossing 
point $x^0_i=m_i/2\mu^2$, the zero mode of the five-dimensional
fermion $\Psi_i(x_\mu,x_5)$ acquires a gaussian wave function 
of typical width $\mu^{-1}$ and centered at $x^0_i$ along the $x_5$
direction: 
\begin{equation}
\Psi^{(0)}_i(x_\mu,x_5)=Ae^{-\mu^2(x_5-x^0_i)^2}\psi_i(x_\mu),
\end{equation}
$\psi_i(x_\mu)$ being a four-dimensional fermion field and 
$A=(2\mu^2/\pi)^{1/4}$ the normalization factor.

Within such a situation, the effective four-dimensional Yukawa
couplings between the five-dimensional SM Higgs boson $H$ and 
zero mode fermions, obtained by integration on $x_5$ over the
wall width $L$:
\begin{eqnarray}
{\cal S}_{Yukawa} & = & \int d^5x \sqrt L \kappa 
H(x_\mu,x_5) \bar \Psi^{(0)}_i(x_\mu,x_5) \Psi^{(0)}_j(x_\mu,x_5) \cr
& = & \int d^4x \lambda_{ij} h(x_\mu) \bar \psi_i(x_\mu) \psi_j(x_\mu),
\label{eq:Yukawa}
\end{eqnarray}
are modulated by the following coupling constants,
\begin{equation}
\lambda_{ij}=\int dx_5 \kappa A^2 
e^{-\mu^2(x_5-x^0_i)^2} e^{-\mu^2(x_5-x^0_j)^2}
=\kappa e^{-{\mu^2\over 2}(x^0_i-x^0_j)^2}.
\label{lambdaij}
\end{equation}
In this context, it can be considered as natural to have a 
dimensionless Yukawa coupling 
constant $\kappa$ which is universal\footnote{The universality applies
here to both the flavor and nature (neutrino, charged lepton, up quark
or down quark) of the particles.} and approximately equal to unity 
at the electroweak scale, and to generate entirely the flavor structure 
of effective Yukawa couplings $\lambda_{ij}$ by field localization 
effects through the exponential suppression factor of 
Eq.(\ref{lambdaij}).

\subsection{Application to the See-saw Model}
\label{seesaw}

Let us apply the AS scenario to the lepton sector.
The SM charged lepton mass hierarchy can effectively be interpreted 
by means of field localization. Indeed, if the zero modes for the 
five-dimensional fields of charged lepton doublets (singlets) under 
$SU(2)_L$ are localized at the positions $l_i$ ($e_j$) along the
wall width, then 
the effective four-dimensional Dirac mass terms can be written as,
\begin{equation}
{\cal L}^{(l^{\pm})}_{Mass}=m^{l^{\pm}}_{ij} \bar e^c_{Li} e^c_{Rj}+h.c.,
\label{LAGmlpm}
\end{equation}
where $e_{Li}$ ($e_{Ri}$) denotes the four-dimensional field of the
left(right)-handed charged lepton and the mass matrix reads as
(see Eq.(\ref{lambdaij})),
\begin{equation}
m^{l^{\pm}}_{ij}= \rho e^{-{\mu^2\over 2}(l_i-e_j)^2},
\ \mbox{with} \ \rho=\kappa \langle h \rangle,
\label{mlpm}
\end{equation}
$\langle h \rangle$ being the vacuum expectation value of the SM Higgs boson.

We now turn to the neutrino sector, assuming that the neutrino
masses result from a see-saw mechanism. Let us recall the basics
of the see-saw model. In this model, a right-handed neutrino $N_R$,
which is a Majorana particle, 
is added to the SM. Then the Lagrangian must contain all the 
additional mass terms involving $N_R$ consistent with the SM gauge 
invariance, which are the following,
\begin{equation}
{\cal L}_{See-saw}=m^{\nu}_{Dij} \bar \nu^c_{Li} N^c_{Rj}
+{1 \over 2} M_{ij} \bar N_{Ri} N^c_{Rj}+h.c.,
\label{Lseesaw}
\end{equation} 
where $\nu_{Li}$ denotes the left-handed neutrino of the SM.
The first term of Eq.(\ref{Lseesaw}) represents a Dirac mass issued from
the spontaneous electroweak symmetry breaking, whereas the second 
term constitutes a Majorana mass originating from a physics underlying 
the SM. This difference of origins allows the Majorana masses to be much 
larger than the Dirac masses, a feature that is required by the see-saw 
mechanism. Under this assumption and after a unitary transformation,
the Lagrangian (\ref{Lseesaw}) can be rewritten in the mass basis as,
\begin{equation}
{\cal L}^{(\nu)}_{Mass} = {1 \over 2} m^{\nu}_{ij} \bar \nu^c_{Li} \nu_{Lj}
+{1 \over 2} M_{ij} \bar N_{Ri} N^c_{Rj}+h.c.,
\label{LseesawMass}
\end{equation}      
where the Majorana mass matrix $m^{\nu}$ is given by the famous 
see-saw formula \cite{Wett}:
\begin{equation}
m^{\nu} \simeq - m^{\nu}_D M^{-1} m^{\nu \ T}_D.
\label{SeesawFormula}
\end{equation}
The flavor structure of the Dirac and Majorana mass matrices
in Eq.(\ref{Lseesaw}), and thus of the neutrino mass matrix 
given by the see-saw formula 
(\ref{SeesawFormula}), can be explained by an AS model. 
For that purpose, the zero modes of the five-dimensional fields 
of both left and right-handed neutrinos must be localized at 
different positions along the wall width. Indeed, in this case,
the Dirac mass matrix of Eq.(\ref{Lseesaw}) reads as, 
\begin{equation}
m^{\nu}_{Dij}= \rho e^{-{\mu^2\over 2}(l_i-N_j)^2},
\label{mnuD}
\end{equation}
exactly like the Dirac mass matrix of charged leptons
(see Eq.(\ref{mlpm})). The parameter $N_j$ in Eq.(\ref{mnuD}) 
is the position of the right-handed neutrino.
Notice that the left-handed charged lepton and left-handed neutrino 
are confined at the same point $l_i$ (see Eq.(\ref{mlpm}) and 
Eq.(\ref{mnuD})). It is due to the fact that the whole
$SU(2)_L$ doublet of leptons is stuck at the point $l_i$.   
Furthermore, in a context of field localization, the
Majorana mass matrix of Eq.(\ref{Lseesaw}) is given by,
\begin{equation}
M_{ij}= M_R e^{-{\mu^2\over 2}(N_i-N_j)^2}.
\label{M}
\end{equation}
In analogy with the Dirac mass matrices of Eq.(\ref{mlpm}) and 
Eq.(\ref{mnuD}), We have assumed 
a common mass scale factor ($M_R$) in the Majorana mass matrix
(\ref{M}). In this way, the flavor structure of mass matrix 
$M_{ij}$ is completely dictated by field displacement effects. 
To obtain the see-saw formula within the AS framework, 
we replace in Eq.(\ref{SeesawFormula}) the Dirac and Majorana mass
matrices by their expression respectively in Eq.(\ref{mnuD}) and 
Eq.(\ref{M}). This leads to the result:
\begin{equation}
m^{\nu}_{ij} \simeq - {\rho^2 \over M_R}
e^{-{\mu^2\over 2}(l_i-N_a)^2}
\bigg [ e^{-{\mu^2\over 2}(N_{a'}-N_{b'})^2} \bigg ] _{ab}^{-1} 
e^{-{\mu^2\over 2}(N_b-l_j)^2},
\label{SeesawFormulaAS}
\end{equation} 
where there is an implicit sum over the $a$ and $b$ indices
and the exponent $-1$ must be taken in the sense of the inverse 
of a matrix. The effective mass matrix of Eq.(\ref{SeesawFormulaAS}) 
is the neutrino mass matrix we will study. Its expression is given
explicitly in Appendix \ref{ExMatrix2F}, for the case of two lepton 
flavors ($\{i,j\}=2,3$).

\subsection{Energy Scales}
\label{subsec:energy}

The three characteristic energy scales of the AS scenario are
the fundamental scale (in a five-dimensional space-time) $M_\star^{5D}$ 
and the energy scales $\mu$ and $L^{-1}$ introduced in Section 
\ref{subsec:AS}.
\\ Let us describe the conditions that these energy scales must fulfilled.
First, $\mu$ must be smaller than the fundamental energy scale. Moreover,
the four-dimensional effective top quark Yukawa coupling has to remain 
perturbative up to the fundamental scale. These two conditions imply the
following scale relation, 
\begin{equation}
\mu < M_\star^{5D} \lesssim 1000L^{-1}.
\label{MassRelationI}
\end{equation}
Besides, for the AS mechanism to make sense, it is necessary that
the wall thickness is larger than the typical width of gaussian wave 
functions $\mu^{-1}$. Furthermore, if 
a field-theoretic description is to work throughout the
domain wall, one must have $\Phi_i(L/2) \sim \mu^2 L < M_\star^{5D}$.
These two new conditions together with Eq.(\ref{MassRelationI}) lead to,
\begin{equation}
\mu^{-1} \lesssim L \lesssim 30\mu^{-1}.
\label{MassRelationII}
\end{equation}

Now, we discuss the value of typical Majorana mass scale $M_R$
characteristic of the see-saw model (see Section \ref{seesaw}), 
within an AS framework. The inequalities (\ref{MassRelationI}) and 
(\ref{MassRelationII}), which summarize the conditions on 
the three energy scales of AS scenario, only constrain ratios of scales
and not the scales themselves. Therefore, we can assume that the
fundamental scale $M_\star^{5D}$ possibly reaches much larger values  
than the electroweak scale (as it was done in \cite{GUT1} for instance). 
Our motivation is to allow the free parameter 
$M_R$ to run over a wide range of orders of magnitude, in order to study
the different modes of the see-saw mechanism. Since we want to 
suppose a large fundamental scale, let us take it higher than 
the GUT scale: $M_\star^{5D} > M_{GUT} \simeq 10^{16}  {\rm GeV}$. In this way,   
the scale of physics beyond the SM, $\Lambda$, which enters the 
non-renormalizable operators of type $(ql)^\dagger(u^cd^c)/\Lambda^2$, 
mediating proton decay, can be assumed 
larger than the GUT scale. This guarantees that the experimental limit on 
proton lifetime, namely $\tau_{proton} \gtrsim 3 \ 10^{33} \ years$, 
is well verified. Besides, for $M_\star^{5D} \gtrsim 10^{16} {\rm GeV}$,
it is clear from Eq.(\ref{MassRelationI}) that the most severe 
experimental bound on the wall width \cite{Delg},
\begin{equation} 
L^{-1} \gtrsim 100 {\rm TeV},  
\label{FCNCboundKK}
\end{equation}
which comes from considering flavor changing neutral currents (FCNC) 
mediated by Kaluza-Klein excitations of gauge bosons, is respected.

Let us specify here our motivations for considering a large fundamental 
mass scale $M_\star^{5D}$ (at the end of our analysis, we will discuss
the implications of a low $M_\star^{5D}$ scale hypothesis).
\\ First, having an high fundamental scale allows to study the case of 
$M_R$ values much larger than
the electroweak scale $\rho$ (see Eq.(\ref{mlpm})). In this case, the 
neutrino mass suppression relatively to $\rho$ 
(see Eq.(\ref{SeesawFormulaAS})) comes more from the factor $\rho / M_R$
(due to the see-saw mechanism) than from the exponential factors (due to
the field localization effects). Such a situation (see end of  
Section \ref{subsec:disres2F}) is interesting since it 
gives rise to leptonic field positions significantly closer to each other
than in the case of neutrino Dirac masses (see Eq.(\ref{mnuD})) where the 
neutrino mass reduction only originates from localization effects.
\\ Secondly, the proton stability can then be ensured by the existence 
of large energy scales ($\Lambda \gtrsim 10^{16} {\rm GeV}$). 
Hence, no suppression of the coupling constants, 
associated to non-renormalizable operators violating both lepton and baryon
numbers, is required from confinement of quarks and 
leptons at far positions in the extra dimension (as proposed by the 
authors of \cite{AS}). 
\\ Similarly, the experimental constraints on FCNC processes are then respected
thanks to the presence of large mass scales (see Eq.(\ref{FCNCboundKK})). 
Therefore, no suppression factors are required from constraining the 
distances between field positions of each quark or lepton generation 
(as suggested in \cite{Bare} for suppressing the rates of FCNC decays: 
$l^\pm_i \to l^\pm_j \gamma$ and $l^\pm_i \to l^\pm_j l^\pm_j l^\mp_j$ 
[$i \neq j$]).     
\\ In summary, choosing an high fundamental scale $M_\star^{5D}$ permits
typically to have closer field positions along the wall. This proximity 
of the fields has two main interests. First, it is in favor of the 
naturality of the AS scenario, as explained in Section \ref{intro}. 
Secondly, it makes the condition on wall thickness $L \lesssim 30\mu^{-1}$ 
(see Eq.(\ref{MassRelationII})), due to considerations on perturbativity, 
easier to fulfill, when one is constructing realistic AS models as
we are going to do in the following.

\section{Realistic AS Models with 2 Flavors}
\label{sec:2flavors}

In this part, we search for the configurations of leptonic field localizations 
in the domain wall which reproduce all the present experimental data,
within the context of AS scenario. 
More precisely, we try to find the values of field positions $e_i,l_j,N_k$
and Majorana mass scale $M_R$ (see Section \ref{seesaw}) being consistent with
the known constraints on both neutrinos and charged leptons. As already said,
we consider the situation where neutrinos acquire masses through the see-saw 
mechanism.

For a better understanding of observable quantity dependence on fundamental
parameters $e_i,l_j,N_k$ and $M_R$, we first treat the case of 2 lepton 
flavors in this part and then the realistic case of 3 flavors in Section
\ref{sec:3flavors}. Hence, in this part, the flavor indices of $e_i,l_j$
and $N_k$ run over the last two families: $\{i,j,k\}=2,3$ (the flavor 
indices hold in the flavor basis).

Besides, in order to impose the experimental constraints progressively, and
thus to restrict our analysis to the relevant regions of theoretical 
parameter space $\{e_i,l_j,N_k,M_R\}$, we proceed through a three step 
approach: we first treat the charged lepton masses, then the leptonic
mixing angles (which are constrained by neutrino oscillation observations) 
and finally the neutrino masses.

\subsection{Charged Lepton Masses}
\label{subsec:CLmasses2F}

\subsubsection{Notations and conventions}

The physical charged lepton masses are derived from a bi-unitary transformation
of the mass matrix, namely $m^{l^{\pm}}_{ij}$ (see Eq.(\ref{mlpm})):
\begin{equation}
m^{l^{\pm}} =
U_{lL}^\dagger
\left ( \begin{array}{cc}
m_{\mu^\pm} & 0  \\ 
0 & m_{\tau^\pm} 
\end{array} \right )
U_{lR},
\label{TRANSmlpm}
\end{equation}
where $L/R$ corresponds to charged lepton chirality (see Eq.(\ref{LAGmlpm})).  
A useful method for computing the charged lepton masses is to diagonalize 
the hermitian square of mass matrix:
\begin{equation}
m^{l^{\pm}} m^{l^{\pm} \ \dagger} =
U_{lL}^\dagger
\left ( \begin{array}{cc}
m^2_{\mu^\pm} & 0  \\ 
0 & m^2_{\tau^\pm} 
\end{array} \right )
U_{lL}.
\label{DIAGmlpm}
\end{equation}
The unitary matrix $U_{lL}$ can be parameterized by a mixing angle $\theta_l$ 
like,
\begin{equation}
U_{lL} =
\left ( \begin{array}{cc}
\cos \theta_l & \sin \theta_l \\ 
- \sin \theta_l & \cos \theta_l
\end{array} \right ).
\label{UlL}
\end{equation}

\subsubsection{Solutions}

Which configurations of the field positions $l_i$ and $e_j$ give rise to a 
mass matrix $m^{l^{\pm}}_{ij}$ (see Eq.(\ref{mlpm})) in agreement with the 
measured values of physical charged lepton masses (see Eq.(\ref{DIAGmlpm})) ? 
The configurations of this kind corresponding to a minimum fine-tuning of
the parameters $l_i$ and $e_j$ are associated to the following textures of 
charged lepton mass matrices,  
\begin{equation}
m^{l^{\pm}}_{I} \simeq
\left ( \begin{array}{cc}
m_\alpha & 0  \\ 
0 & m_\beta 
\end{array} \right ), \
m^{l^{\pm}}_{II} \simeq
\left ( \begin{array}{cc}
0 & m_\alpha  \\ 
m_\beta & 0
\end{array} \right ),
\label{textures2F}
\end{equation}
$m_{\alpha,\beta}$ being the experimental values of charged lepton masses:
$\{m_\alpha,m_\beta\}=m^{exp}_{\mu^\pm},m^{exp}_{\tau^\pm}$.
This result will be shown numerically at the end of this Section, and it can 
be understood as follows. First, the textures of Eq.(\ref{textures2F})
reproduce well the measured charged lepton masses, as it is clear in the
example of the diagonal case:
\begin{equation}
m^{l^{\pm}}_{I} \simeq
\left ( \begin{array}{cc}
m^{exp}_{\mu^\pm} & 0  \\ 
0 & m^{exp}_{\tau^\pm}
\end{array} \right ) \ \Rightarrow \ 
m^{l^{\pm}}_{I} m^{l^{\pm} \ \dagger}_{I} \simeq 
U_{lL}^\dagger 
\left ( \begin{array}{cc} 
m^{exp \ 2}_{\mu^\pm} & 0  \\ 
0 & m^{exp \ 2}_{\tau^\pm} 
\end{array} \right )
U_{lL}, \ U_{lL} \simeq {\bf 1}_{2 \times 2}.
\label{diagcase}
\end{equation}
Secondly, the textures in Eq.(\ref{textures2F}) correspond to a
minimal fine-tuning of $l_i$ and $e_j$ since they impose only
one specific condition per parameter. For instance, in the diagonal case,
the only mass relation that the parameter $l_2$ has to verify is
(see Eq.(\ref{mlpm}) and Eq.(\ref{diagcase})), 
\begin{equation}
m^{exp}_{\mu^\pm} = \rho e^{-{\mu^2\over 2}(l_2-e_2)^2}.
\end{equation}
We also see on this example 
that the choice in the sign of quantity $l_i-e_j$ entering Eq.(\ref{mlpm}) is 
arbitrary, leading thus to different types of acceptable field position 
configurations.

All the observable quantities (masses,\dots) are invariant under some 
trivial transformations of the localization configurations. For instance,  
the physical quantities are left invariant by a simultaneous translation of 
all the field positions along the domain wall. This is due to the fact
that the mass matrices involve only some differences of the positions
(see Eq.(\ref{mlpm})). Therefore, in order to not consider localization 
configurations which are physically equivalent to each other, we fix the 
value for one of the positions: $e_2=0$. In the example of diagonal
charged lepton mass matrix described above, this choice leads to the 
relation, 
\begin{equation}
m^{exp}_{\mu^\pm} = \rho e^{-{\mu^2\over 2}l_2^2},
\end{equation}
which gives the absolute value of parameter $l_2$. 
\\ Similarly, there
is an invariance under the symmetry $X_i\to-X_i$, $X_i$ representing all 
the field positions. This one comes from the fact that only some squares 
of position differences enter the mass matrices (see Eq.(\ref{mlpm})).
Hence, for the same reason as before, we fix the sign for one of the
$l_i$ parameters. In the example of diagonal mass matrix, we would fix
the sign of $l_2$ so that this parameter would be completely determined:
\begin{equation}
l_2 = \mu^{-1} \sqrt{ -2\ln(m^{exp}_{\mu^\pm}/\rho) }.
\end{equation}
In summary, eliminating physically equivalent localization 
configurations allows to determine one of the $l_i$ parameters, 
if one considers the charged lepton textures of Eq.(\ref{textures2F}).

\subsubsection{Mass uncertainty}

Although the charged lepton masses are known up to an high precision, it is
more reasonable to consider an existing significant uncertainty on the 
measured values of 
those masses. The three reasons are the following. First, the goal of our 
analysis is not to find the values of acceptable field positions with an high 
accuracy. Secondly, a non-negligible error on observable quantities must 
be introduced if the fine-tuning of fundamental parameters is to be discussed. 
The third reason has to do with the fact that the running of lepton masses 
with the energy scale must be taken into account in the analysis. 
In order to study 
lepton mass hierarchies which are not affected by renormalization effects 
(for simplicity), we consider all the mass values at a common energy scale 
of the order of electroweak scale, namely the top quark mass $m_{top}$. 
Therefore, the theoretical predictions for lepton mass values are computed 
with (see Eq.(\ref{mlpm})), 
\begin{equation}
\rho=\rho(m_{top})=\kappa(m_{top})\langle h \rangle=1.5 m_{top}(m_{top}),
\label{rhomtop}
\end{equation}
where the mass of the top quark evaluated at its own mass scale
in the $\overline{MS}$ 
scheme is $m_{top}(m_{top})=166 \pm 5 {\rm GeV}$, following the studies of 
\cite{Mira,Bare}. The choice of $\rho(m_{top})$ value is motivated by the 
facts that the AS mechanism works for $\rho \geq m_{top}$ and that
considerations on naturalness and perturbativity lead to a coupling
constant $\kappa(m_{top})$ close to unity (see Section \ref{subsec:AS} and 
discussion in \cite{Mira}). This choice does not affect our results and 
predictions as we will discuss later (Section \ref{subsec:rho}). 
Since the experimental values of lepton masses 
and their theoretical predictions must be compared at an identical energy 
scale, the measured lepton masses have also to be taken at the top mass 
scale. Nevertheless, the effect on the charged lepton masses of running 
from the pole mass scale up to the top mass scale is only of a few percents 
\cite{Mira}. 
Hence, for the experimental values of charged lepton masses, we take the
pole masses \cite{PDG} and we assume an uncertainty of $5 \%$: 
$\delta m^{exp}_{l^\pm}/m^{exp}_{l^\pm}=0.05$.

In case of a significant uncertainty on the measured values of 
charged lepton masses, there is a possible deviation from the 
textures considered in Eq.(\ref{textures2F}) which reproduce the 
correct masses. Let us consider once more the 
example of diagonal case: the existence of uncertainties on experimental 
charged lepton masses allows a continuous variation from the realistic 
texture in Eq.(\ref{diagcase}):
\begin{equation}
m^{l^{\pm}}_{I} =
\left ( \begin{array}{cc}
m^{exp}_{\mu^\pm}+\epsilon_1 & \epsilon_2 \\ 
\epsilon_3 & m^{exp}_{\tau^\pm}+\epsilon_4
\end{array} \right ) \ \Rightarrow \ 
m^{l^{\pm}}_{I} m^{l^{\pm} \ \dagger}_{I} = 
U_{lL}^\dagger 
\left ( \begin{array}{cc} 
m^{exp \ 2}_{\mu^\pm}+\delta_1 & \delta_2  \\ 
\delta_3 & m^{exp \ 2}_{\tau^\pm}+\delta_4 
\end{array} \right )
U_{lL},
\label{diagcaseDM}
\end{equation}
$\epsilon_i$ and $\delta_i$ representing mass variations. 
The localization configurations associated to texture (\ref{diagcaseDM}),
which give rise to a minimum fine-tuning of parameters $l_i$ and $e_j$, correspond to 
$\delta_2,\delta_3 \ll m^{exp \ 2}_{\mu^\pm},m^{exp \ 2}_{\tau^\pm}$ with 
$\delta_1 \in [-\delta m^{exp \ 2}_{\mu^\pm},+\delta m^{exp \ 2}_{\mu^\pm}]$ 
and
$\delta_4 \in [-\delta m^{exp \ 2}_{\tau^\pm},+\delta m^{exp \ 2}_{\tau^\pm}]$,
leading to $\theta_l \approx 0$ (see Eq.(\ref{UlL})), as it will be shown 
in Section \ref{subsec:angles2F}.
The reason is that the presence of non-negligible contributions to 
$m^{l^{\pm}}_{I} m^{l^{\pm} \ \dagger}_{I}$ from 
$\delta_{2,3}=f(\epsilon_i)=g(l_i,e_j)$ ($f$ and $g$ being certain functions) 
would give rise to new specific relations involving $l_i$ and $e_j$, 
and would thus increase their fine-tuning.
\\ In conclusion, even with significant uncertainties on measured charged 
lepton masses, the less fine-tuned solutions of field positions $l_i$ and $e_j$
associated to textures in Eq.(\ref{textures2F}) lead to: $\theta_l \approx 0$ 
for $[m_\alpha=m^{exp}_{\mu^\pm}; m_\beta=m^{exp}_{\tau^\pm}]$ and 
$\theta_l \approx \pi/2$ for $[m_\alpha=m^{exp}_{\tau^\pm};
m_\beta=m^{exp}_{\mu^\pm}]$. This result can be understood like this: 
requiring a given large mixing in the charged lepton sector constitutes 
an additional specific condition on $l_i$ and $e_j$, and thus tends to 
increase their fine-tuning.

\subsubsection{Scan}

By performing a simultaneous scan on the parameters $l_2,l_3$ and $e_3$ 
($e_2=0$ as explained before) with a step of $10^{-2}$ 
in the ranges $l_2,l_3\in[-15\mu^{-1},15\mu^{-1}]$ and 
$e_3\in[-20\mu^{-1},20\mu^{-1}]$ (this choice is motivated by 
Eq.(\ref{MassRelationII})), we find that all the localization configurations
reproducing the wanted charged lepton masses correspond to the following 
textures,
\begin{equation}
m^{l^{\pm}}_{I} \simeq
\left ( \begin{array}{cc}
m_\alpha & \epsilon_1  \\ 
\epsilon_2 & m_\beta 
\end{array} \right ), \
m^{l^{\pm}}_{II} \simeq
\left ( \begin{array}{cc}
\epsilon_1 & m_\alpha  \\ 
m_\beta & \epsilon_2
\end{array} \right ),
\label{finaltextures2F}
\end{equation}
$m_{\alpha,\beta}$ being defined as in Eq.(\ref{textures2F}), with 
$\theta_l \approx 0$ or $\theta_l \approx \pi/2$. 
This result shows that the typically less fine-tuned 
solutions\footnote{In the sense that these solutions are obtained via a 
coarse scan, with a step of $10^{-2}$.}, for
field positions $l_i$ and $e_j$, correspond indeed to continuous deviations
(because of the significant mass uncertainty) 
from the textures (\ref{textures2F}) with $\theta_l \approx 0$ or $\pi/2$.
For instance, one class of solutions, reproducing the correct charged lepton 
masses, that we find via the scan described previously reads as, 
\begin{eqnarray}
\bigg \{ 
e_2=0; \  
e_3 \simeq l_3+\mu^{-1} \sqrt{ -2\ln(m^{exp}_{\tau^\pm}/\rho) }; \
l_2 \simeq \mu^{-1} \sqrt{ -2\ln(m^{exp}_{\mu^\pm}/\rho) }; \cr
l_3 \in [-15,-3.61]\mbox{U}[4.41,15] \ (\mbox{in units of} \ \mu^{-1})
\bigg \}.
\label{eq:exsol2F}
\end{eqnarray}
This type of solutions corresponds (see Eq.(\ref{mlpm})) to the texture 
$m^{l^{\pm}}_{I}$ of Eq.(\ref{finaltextures2F}) with 
$m_\alpha=m^{exp}_{\mu^\pm}$, $m_\beta=m^{exp}_{\tau^\pm}$ and $e_3-l_3>0$
(the sign of $l_2$ being fixed in order to eliminate the equivalent 
solutions obtained by the action of symmetry $X_i\to-X_i$).

\subsubsection{Conclusion}

In summary, the less fine-tuned solutions for field positions $l_i$ and $e_j$, 
reproducing the measured charged lepton masses, are divided into 8 types of
solutions corresponding to: 
2 kinds of matrix texture (see Eq.(\ref{finaltextures2F})) 
$\times$ 2 mass permutations ($[m_\alpha=m^{exp}_{\mu^\pm}; 
m_\beta=m^{exp}_{\tau^\pm}]$ or $[m_\alpha=m^{exp}_{\tau^\pm};
m_\beta=m^{exp}_{\mu^\pm}]$) 
$\times$ 2 opposite signs (for one of the differences $l_i-e_j$). 
Those classes of solutions are such that the charged lepton mixing
angle (see Eq.(\ref{UlL})) is given by $\theta_l \approx 0$
for $[m_\alpha=m^{exp}_{\mu^\pm}; m_\beta=m^{exp}_{\tau^\pm}]$ and
$\theta_l \approx \pi/2$ for $[m_\alpha=m^{exp}_{\tau^\pm};
m_\beta=m^{exp}_{\mu^\pm}]$.

\subsection{Mixing Angles}
\label{subsec:angles2F}

\subsubsection{Notations and conventions}

Within the see-saw model, the left-handed neutrinos acquire a mass of
Majorana type (see Eq.(\ref{LseesawMass})) so that the physical masses 
can be directly obtained from a diagonalization of neutrino mass matrix 
$m^{\nu}_{ij}$ (see Eq.(\ref{SeesawFormulaAS})):
\begin{equation}
m^{\nu} =
U_{\nu L}^t
\left ( \begin{array}{cc}
m_{\nu_2} & 0  \\ 
0 & m_{\nu_3} 
\end{array} \right )
U_{\nu L},
\label{TRANSmnu}
\end{equation}
where $U_{\nu L}$ satisfies $U_{\nu L}^t=U_{\nu L}^\dagger=U_{\nu L}^{-1}$ 
(since within our whole study, we assume the absence of any CP violation phase 
in lepton mass matrices\footnote{The presence of non-vanishing complex phases 
would not affect our results, except possibly our predictions on the effective 
neutrino mass constrained by neutrinoless double $\beta$ decay experiments.})
and can be parameterized by a mixing angle $\theta_\nu$ as in Eq.(\ref{UlL}). 
In our analysis, we consider only the case of normal hierarchy for 
neutrino mass eigenvalues, namely $m_{\nu_2}<m_{\nu_3}$ for 2 lepton 
flavors. Nevertheless, in the discussion on our results, we will 
describe the effect of having instead another neutrino mass hierarchy.    
In preparation of future discussions, we introduce the quantities 
$V_i=V_i(l_j,N_k)$ which are the eigenvalues of the dimensionless matrix 
$(M_R / \rho^2)m^{\nu}_{ij}$ (see Eq.(\ref{SeesawFormulaAS})). This 
definition leads to the following expression for the left-handed neutrino 
mass eigenvalues, 
\begin{equation}
m_{\nu_i} = {\rho^2 \over M_R} V_i(l_j,N_k). 
\label{eq:useful}
\end{equation}
With the above conventions, the lepton mixing matrix $U_{MNS}$ 
\cite{PMNS} appearing in the leptonic
charged current, ${\cal L}_{CC}= - (g / \sqrt 2) 
\bar e^m_L \gamma^\mu U_{MNS} \nu^m_L W^-_\mu + h.c.$
(the exponents $m$ stand for mass basis), reads as,
\begin{equation}
U_{MNS} =U_{lL}U_{\nu L}^\dagger.
\label{UmnsI}
\end{equation} 
Hence, the unitary matrix $U_{MNS}$ can be parameterized as
in Eq.(\ref{UlL}) with a mixing angle $\theta_{23}$ given by: 
\begin{equation}
\theta_{23}=\theta_l-\theta_\nu. 
\label{theta23A}
\end{equation}

\subsubsection{Solutions}

Here, we search for the less fine-tuned solutions of field positions 
$e_i,l_j$ and $N_k$, determining both $m^{l^{\pm}}_{ij}$ (see Eq.(\ref{mlpm})) 
and $m^{\nu}_{ij}$ (see Eq.(\ref{SeesawFormulaAS})), which lead to 
transformation matrices $U_{lL}$ (see Eq.(\ref{DIAGmlpm})) and $U_{\nu L}$ 
(see Eq.(\ref{TRANSmnu})), and thus to a mixing matrix $U_{MNS}$ (see 
Eq.(\ref{UmnsI})), compatible with experimental data on neutrino oscillation 
physics.
For that purpose, we use results obtained through a coarse scan over the 
parameter space $\{N_2,N_3\}$, for values of $e_2,e_3,l_2$ and $l_3$
which belong to the less fine-tuned solutions reproducing charged lepton 
masses described in Section \ref{subsec:CLmasses2F}. Indeed, we consider
simultaneously the charged lepton and neutrino sectors which are related
through the mixing matrix $U_{MNS}$ (see Eq.(\ref{UmnsI})) and field 
positions $l_i$ (see Eq.(\ref{mlpm}) and Eq.(\ref{SeesawFormulaAS})).

\begin{figure}[!t]
\begin{center}
\begin{tabular}{cc}
\psfrag{N2}[c][c][1]{$N_2$}
\psfrag{N3}[c][c][1]{$N_3$}
\includegraphics[width=0.5\textwidth,height=4cm]{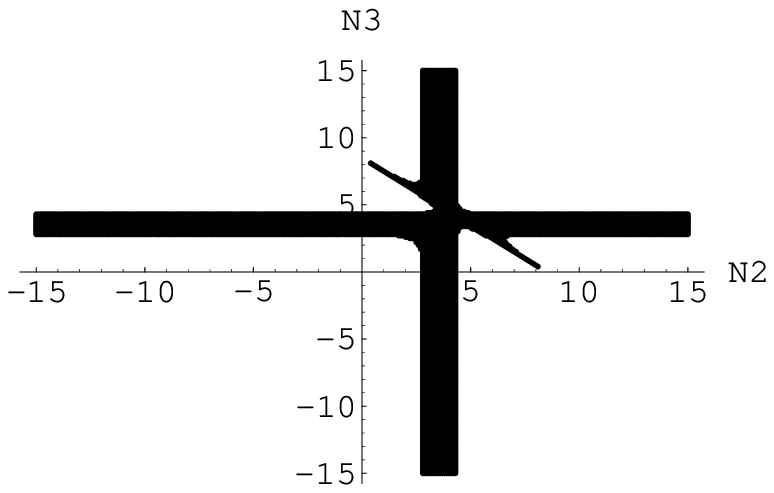} &
\psfrag{N2}[c][c][1]{$N_2$}
\psfrag{N3}[c][c][1]{$N_3$}
\includegraphics[width=0.5\textwidth,height=4cm]{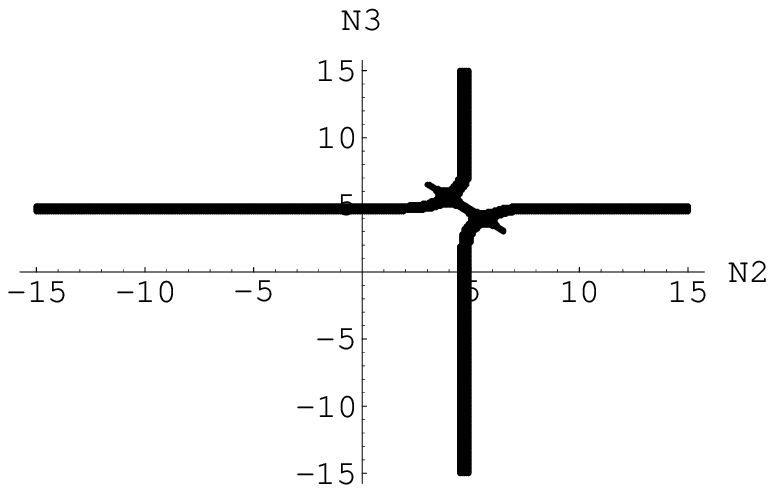} \\
$[a]$ & $[b]$ \\
 & \\
\psfrag{N2}[c][c][1]{$N_2$}
\psfrag{N3}[c][c][1]{$N_3$}
\includegraphics[width=0.5\textwidth,height=4cm]{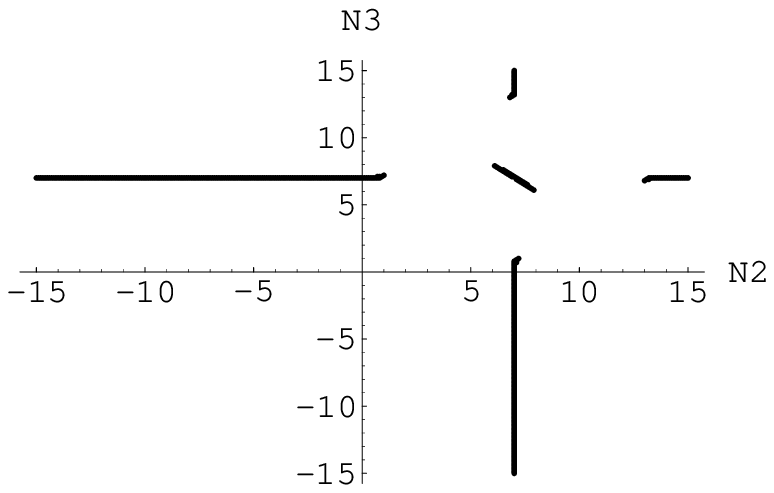} &
\psfrag{N2}[c][c][1]{$N_2$}
\psfrag{N3}[c][c][1]{$N_3$}
\includegraphics[width=0.5\textwidth,height=4cm]{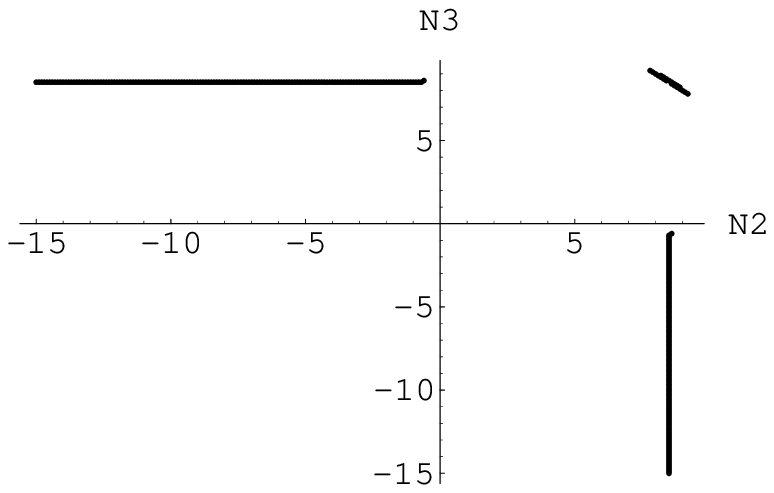} \\
$[c]$ & $[d]$
\end{tabular}
\caption{Domains of the plane $\{N_2,N_3\}$ (units of $\mu^{-1}$) 
corresponding to $0.5<\tan^2\theta_{23}<2.5$ (experimental range 
at $99 \%  \ C.L.$ \cite{data}),
for $\{ e_2=0; e_3=l_3+3.14\mu^{-1}; l_2=3.94\mu^{-1} \}$ and
$l_3=4.5\mu^{-1}$ [a], $l_3=5.6\mu^{-1}$ [b], $l_3=10.1\mu^{-1}$ [c] or
$l_3=13.1\mu^{-1}$ [d] (all these 4 points in space $\{e_2,e_3,l_2,l_3\}$
belong to the class of solutions (\ref{eq:exsol2F})
fitting the values of $m^{exp}_{e^\pm,\mu^\pm,\tau^\pm}$).
Those regions have been derived from the scan described in text.}
\label{fig:cross}
\end{center}
\end{figure}

For instance, let us consider the class of solutions for charged 
lepton masses given in Eq.(\ref{eq:exsol2F}), which is one of the 8 types 
of solutions described at the end of 
Section \ref{subsec:CLmasses2F}. For some different values of 
$l_3$ included in the range of Eq.(\ref{eq:exsol2F}), we find, via a scan 
over parameter space $\{N_2,N_3\}$, the regions of $\{N_2,N_3\}$ which 
give rise to a mixing angle $\theta_{23}$ 
consistent with the experimental constraint from observation of 
atmospheric neutrino oscillations ($\nu_\mu \leftrightarrow \nu_\tau$): 
$0.5<\tan^2\theta_{23}<2.5$ at $99 \%  \ C.L.$ \cite{data}.
For this scan, the interval explored is $N_2,N_3\in[-15\mu^{-1},15\mu^{-1}]$ 
(as suggested by Eq.(\ref{MassRelationII})) and the step used is $10^{-1}$ 
(the values of $e_2,e_3,l_2,l_3$ have been obtained through the scan 
described at the end of Section \ref{subsec:CLmasses2F}).
The results are presented in Fig.(\ref{fig:cross}).

On the four graphics of Fig.(\ref{fig:cross}), we see clearly that there 
are two distinct kinds of solutions reproducing a mixing angle $\theta_{23}$ 
which fulfills the experimental condition $0.5<\tan^2\theta_{23}<2.5$ 
\cite{data}: while the first type of solutions verifies,
\begin{equation}
N_2 \approx {l_2+l_3 \over 2} \ \mbox{or} \ N_3 \approx {l_2+l_3 \over 2}, 
\label{eq:SOLcross2F}
\end{equation}
the other one is such that,
\begin{equation}
N_3 \approx l_2+l_3-N_2. 
\label{eq:SOLdiag2F}
\end{equation}
In the following, we explain and interpret those two kinds of solutions.

\subsubsection{Interpretation}

{\bf \quad a)} Let us first discuss the type of solutions characterized by 
$N_2 \approx (l_2+l_3)/2$ (see Eq.(\ref{eq:SOLcross2F})). For those
solutions, if $N_3$ is sufficiently far from $N_2,l_2$ and $l_3$
(see Fig.(\ref{fig:cross})[c,d]),
the diagonal elements of neutrino mass matrix $m^{\nu}_{ij}$ (given 
in Eq.(\ref{SeesawFormulaAS})) are approximately equal: 
\begin{equation}
m^{\nu}_{22} \approx m^{\nu}_{33} \approx 
- {\rho^2 \over M_R} e^{-{\mu^2\over 4}(l_2-l_3)^2}.
\label{22eq33}
\end{equation}
Indeed, in this case, the first term, in the expressions of 
$m^{\nu}_{22}$ and $m^{\nu}_{33}$ given by Eq.(\ref{eq:ExMa2F}), 
is dominant compared to other terms.
Now if $l_2$ is close enough to $l_3$, then for any value of $N_3$ 
(see Fig.(\ref{fig:cross})[a,b]), the diagonal elements are also
nearly identical (see Eq.(\ref{eq:ExMa2F})). The same demonstration
can be done for the other situation: $N_3 \approx (l_2+l_3)/2$. 
Therefore, the solutions associated to Eq.(\ref{eq:SOLcross2F}) 
lead to $m^{\nu}_{22} \approx m^{\nu}_{33}$.

\begin{figure}[!t]
\begin{center} 
\begin{tabular}{cc}
\psfrag{nu2}[c][c][1]{$\nu_{L2}$}
\psfrag{nu2c}[c][c][1]{$\nu_{L2}$}
\psfrag{Nu2}[c][ct][1]{$N_{R2}$}
\psfrag{h}[c][c][1]{$\langle h \rangle$}
\psfrag{M22}[c][c][1]{$M_{22}$}
\psfrag{lbda22}[c][c][1]{$\lambda^{\nu}_{22}$}
\includegraphics[width=0.5\textwidth,height=2.5cm]{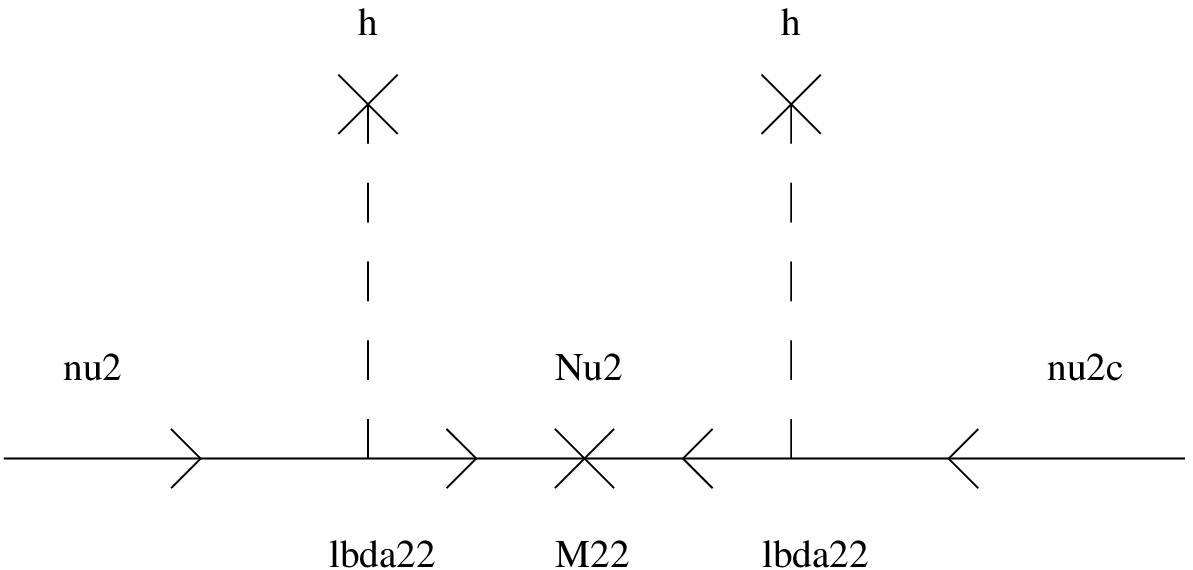} &
\psfrag{nu3}[c][c][1]{$\nu_{L3}$}
\psfrag{nu3c}[c][c][1]{$\nu_{L3}$}
\psfrag{Nu2}[c][ct][1]{$N_{R2}$}
\psfrag{h}[c][c][1]{$\langle h \rangle$}
\psfrag{M22}[c][c][1]{$M_{22}$}
\psfrag{lbda32}[c][c][1]{$\lambda^{\nu}_{32}$}
\includegraphics[width=0.5\textwidth,height=2.5cm]{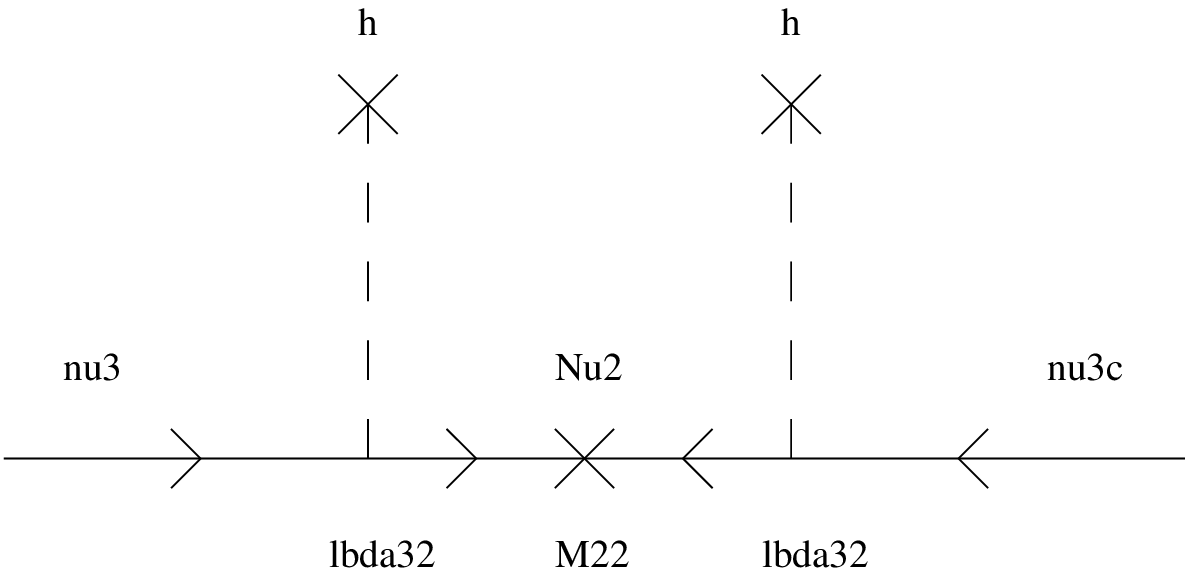} \\
 & \\
$[a]$ & $[b]$
\end{tabular}
\caption{Feynman diagrams of all processes generating the elements
$m^{\nu}_{22}$ [a] and $m^{\nu}_{33}$ [b] of mass matrix 
$m^{\nu}_{ij}$ for left-handed Majorana neutrinos (see 
Eq.(\ref{LseesawMass})), within the see-saw model when 
the neutrino $N_{R3}$ is weakly coupled to other neutrinos 
(only the lepton flavors $\{i,j\}=2,3$ are considered). $h$, 
$N_{Ri}$ and $\nu_{Li}$ stand respectively for the fields 
of SM Higgs boson, right and left-handed Majorana neutrinos. 
$M_{22}$ denotes a diagonal element of Majorana mass matrix 
$M_{ij}$ for right-handed neutrinos (see Eq.(\ref{Lseesaw})). 
$\lambda^{\nu}_{22}$ and $\lambda^{\nu}_{32}$ represent 
Yukawa couplings. Finally, the crosses indicate mass insertions 
or vacuum expectation values, and, the arrows show the flow 
of momentum for the associated fields.}
\label{fig:seesaw}
\end{center}
\end{figure}

This result that solutions associated to Eq.(\ref{eq:SOLcross2F})
give rise to $m^{\nu}_{22} \approx m^{\nu}_{33}$ (diagonal elements
of mass matrix (\ref{SeesawFormulaAS})) can be explained in the
following terms. In the see-saw model, the small Majorana masses 
for left-handed neutrinos ($m^{\nu}_{ij}$ in Lagrangian (\ref{LseesawMass}))
are generated by the exchange of heavy right-handed Majorana
neutrinos. For the solutions of type $N_2 \approx (l_2+l_3)/2$ with 
$N_3$ sufficiently far from $N_2,l_2$ and $l_3$, the right-handed Majorana 
neutrino $N_{R3}$ has a negligible effective coupling to all other neutrinos, 
because of a weak overlap of gaussian wave functions in the extra dimension.
Hence, the Majorana masses $m^{\nu}_{22}$ and $m^{\nu}_{33}$ for left-handed 
neutrinos, respectively $\nu_{L2}$ and $\nu_{L3}$, are generated only by the 
same exchange of the right-handed Majorana neutrino $N_{R2}$, as illustrated 
in Fig.(\ref{fig:seesaw}). The consequence is that the difference between 
$m^{\nu}_{22}$ and $m^{\nu}_{33}$ can only originate from the difference
between Yukawa coupling constants of neutrinos $\lambda^{\nu}_{22}$ and 
$\lambda^{\nu}_{32}$ (see Eq.(\ref{eq:Yukawa}) and Fig.(\ref{fig:seesaw})). 
Now, since we are in the situation where $N_2 \approx (l_2+l_3)/2$, the 
Yukawa coupling constants (see Eq.(\ref{lambdaij}), Eq.(\ref{mlpm}) and
Eq.(\ref{mnuD})),
\begin{equation}
\lambda^{\nu}_{22}=\kappa e^{-{\mu^2\over 2}(l_2-N_2)^2}
\ \mbox{and} \ 
\lambda^{\nu}_{32}=\kappa e^{-{\mu^2\over 2}(l_3-N_2)^2},
\label{eq:lambdanu}
\end{equation}
are approximately equal so that $m^{\nu}_{22} \approx m^{\nu}_{33}$.  
More precisely, for $N_2 \approx (l_2+l_3)/2$, these two mass elements 
are nearly equal to the common expression (see Fig.(\ref{fig:seesaw}),
Eq.(\ref{mlpm}) and Eq.(\ref{eq:lambdanu})):
\begin{equation}
m^{\nu}_{22} \approx m^{\nu}_{33} \approx
- {(\lambda^{\nu}_{32}\langle h \rangle)^2 \over M_{22}} = 
- {\rho^2 \over M_{22}} e^{-\mu^2(l_3-N_2)^2} \approx 
- {\rho^2 \over M_{22}} e^{-{\mu^2\over 4}(l_2-l_3)^2}.
\end{equation}
This relation provides an interpretation of the approximately common value
for $m^{\nu}_{22}$ and $m^{\nu}_{33}$ found in Eq.(\ref{22eq33}), since 
one has $M_{22}=M_R$ (see Eq.(\ref{M})).

Since the type of solutions associated to Eq.(\ref{eq:SOLcross2F}) 
corresponds to $m^{\nu}_{22} \approx m^{\nu}_{33}$, it leads to a quasi 
maximal mixing in the neutrino sector: 
\begin{equation}
\theta_\nu \approx \pi/4.
\label{theta23B}
\end{equation}
As a matter of fact, the mixing angle $\theta_\nu$
parameterizing the orthogonal matrix $U_{\nu L}$, which allows to diagonalize 
(see Eq.(\ref{TRANSmnu})) the real and symmetric neutrino Majorana mass 
matrix $m^{\nu}_{ij}$ of Eq.(\ref{SeesawFormulaAS}), is defined by, 
\begin{equation}
\tan 2 \theta_{\nu}={2 m^{\nu}_{23} \over m^{\nu}_{33} - m^{\nu}_{22}}. 
\label{eq:tg2tetanu}
\end{equation}
From Eq.(\ref{theta23A}) and Eq.(\ref{theta23B}), we deduce that the
solutions (\ref{eq:SOLcross2F}), which reproduce a mixing angle $\theta_{23}$ 
in agreement with the experimental constraint $0.5<\tan^2\theta_{23}<2.5$ 
\cite{data} (or equivalently $\vert \theta_{23} \vert \approx \pi/4$), 
correspond to,  
\begin{equation}
\theta_l \approx 0.
\end{equation}
This means that the solutions in $\{l_i,e_j\}$ for charged lepton masses, 
that we have considered in Fig.(\ref{fig:cross}) (class of solutions 
(\ref{eq:exsol2F})), generate a nearly vanishing mixing in the
charged lepton sector, as we said in Section \ref{subsec:CLmasses2F}.

The width along $N_2$, for domains in $\{N_2,N_3\}$ associated to the 
kind of solutions $N_2 \approx (l_2+l_3)/2$ (see Eq.(\ref{eq:SOLcross2F})), 
is getting smaller as the absolute difference $\vert l_2-l_3 \vert$ 
increases (see Fig.(\ref{fig:cross})). This is due to the fact that,  
when $\vert l_2-l_3 \vert$ increases, $m^{\nu}_{22}$ and $m^{\nu}_{33}$
decrease (see Eq.(\ref{22eq33})) so that adjusting them to an almost 
common value (in order to still have $\theta_\nu \approx \pi/4$) requires 
an higher accuracy on the configurations of positions $l_i$ and $N_j$.
The same argument holds for the solutions of type $N_3 \approx (l_2+l_3)/2$.

{\bf b)} Now, we discuss the other class of solutions generating a mixing 
angle $\theta_{23}$ consistent with experimental data, namely the solutions 
which satisfy $N_3 \approx l_2+l_3-N_2$ (see Eq.(\ref{eq:SOLdiag2F})).
For those solutions, the diagonal elements of neutrino mass matrix 
$m^{\nu}_{ij}$ (given in Eq.(\ref{SeesawFormulaAS})) are also of the 
same order (see Eq.(\ref{eq:ExMa2F})):
\begin{eqnarray}
m^{\nu}_{22} \approx m^{\nu}_{33} \approx
- {\rho^2 \over M_R (1-e^{-\mu^2(l_2+l_3-2N_2)^2})}
(e^{-\mu^2(l_2-N_2)^2}+e^{-\mu^2(l_3-N_2)^2} \cr
-2e^{-{\mu^2\over 2}[(l_2+l_3-2N_2)^2+(l_2-N_2)^2+(l_3-N_2)^2]}).
\label{22eq33bis}
\end{eqnarray}
As before, this almost equality between $m^{\nu}_{22}$ and $m^{\nu}_{33}$ 
can be understood from a diagrammatic point of view, and, it corresponds 
to a quasi maximum mixing in the neutrino sector: $\theta_\nu \approx \pi/4$.
On Fig.(\ref{fig:cross}), these solutions of type $N_2+N_3 \approx l_2+l_3$ 
do not appear in regions where $N_2$ is too far from $N_3$. The reasons 
are that, for this type of solution, the accuracy on $l_i$ and $N_j$ becomes 
higher as $\vert N_2-N_3 \vert$ increases, and, that the results presented 
in Fig.(\ref{fig:cross}) have been derived from a scan on field position
values with a given step.

\subsubsection{Conclusion}

We have shown that the less fine-tuned solutions
in $\{e_i,l_j,N_k\}$, consistent with the experimental values for 
both charged lepton masses and mixing matrix $U_{MNS}$, satisfy 
either the condition $N_2,N_3 \approx (l_2+l_3)/2$ (of  
Eq.(\ref{eq:SOLcross2F})) or $N_2+N_3 \approx l_2+l_3$ (of    
Eq.(\ref{eq:SOLdiag2F})) which lead to a quasi maximal mixing in 
the neutrino sector: $\theta_\nu \approx \pi/4$. Moreover, these 
solutions correspond to the configurations of field positions $e_i$
and $l_j$ described in Section \ref{subsec:CLmasses2F} which give rise
to $\theta_l \approx 0$.

\subsection{Neutrino Masses}
\label{subsec:NuMass}

The neutrino mass squared difference $\Delta m^2_{23}=m^2_{\nu_3}-m^2_{\nu_2}$ 
(see Eq.(\ref{TRANSmnu})) can be written as (see Eq.(\ref{eq:useful})),
\begin{equation}
\Delta m^2_{23}= {\rho^4 \over M_R^2}(V^2_3-V^2_2).
\label{DM23}
\end{equation}
We note that the physical quantity 
$\Delta m^2_{23}$ depends on the parameters $l_i,N_j$ and $M_R$. In 
contrast, by looking at Eq.(\ref{eq:tg2tetanu}) we see that the mixing 
angle of 
neutrino sector $\theta_\nu$, and thus the matrix $U_{MNS}$ (see 
Eq.(\ref{theta23A})), does not depend on $M_R$ 
which enters the neutrino mass matrix $m^{\nu}_{ij}$ via an overall
factor $\rho^2/M_R$ (see Eq.(\ref{SeesawFormulaAS})). The equality 
(\ref{DM23}) can be rewritten as,
\begin{equation}
M_R= {\rho^2 \over \sqrt{\Delta m^2_{23}}} \sqrt{V^2_3-V^2_2}.
\label{DM23bis}
\end{equation}
From this relation, it is clear that if we fix the quantity 
$\Delta m^2_{23}$ to a value $\Delta m^{2 \ exp}_{23}$ contained 
in the range allowed by experimental data, the parameter $M_R$ becomes 
a function of the other parameters $l_i$ and $N_j$. 
Here, we set $\Delta m^2_{23}$ at the value favored by
atmospheric neutrino oscillations: 
$\Delta m^2_{23}=\Delta m^{2 \ exp}_{23}=2.6 \ 10^{-3} {\rm eV}^2$ (best-fit point
obtained by a $\chi^2$ method) \cite{data}, and we discuss the behavior of
$M_R$ dictated by Eq.(\ref{DM23bis}) along the regions of parameter space 
$\{l_i,N_j\}$ shown in Fig.(\ref{fig:cross}). Let us remind that those regions 
are included in domains of space $\{e_i,l_j,N_k\}$ consistent with the 
experimental values for charged lepton masses and leptonic mixing angles.

Let us first consider the regions of Fig.(\ref{fig:cross}) where 
$N_2 \approx (l_2+l_3)/2$ (with $N_3$ significantly far from $N_2$).
There, each element of the matrix $-(M_R / \rho^2)m^{\nu}_{ij}$ is 
approximately equal to $e^{-{\mu^2\over 4}(l_2-l_3)^2}$ (see 
Eq.(\ref{eq:ExMa2F})). This is the reason why, in these regions, the 
value of $M_R$ (given by Eq.(\ref{DM23bis})) remains nearly constant through 
the plane $\{N_2,N_3\}$ whereas it decreases when $\vert l_2-l_3 \vert$ becomes
larger. In those domains, one has indeed (see Fig.(\ref{fig:cross})): 
\begin{eqnarray}
M_R \approx 10^{15} {\rm GeV}  \ \mbox{in [a]},
& 
M_R \approx 10^{15} {\rm GeV}  \ \mbox{in [b]},\nonumber 
\\
M_R \approx 2 \  10^{11} {\rm GeV}  \ \mbox{in [c]},
& 
M_R \approx 2 \  10^6 {\rm GeV}  \ \mbox{in [d]}.\nonumber
\end{eqnarray}
In Fig.(\ref{fig:cross})[a], $l_2-l_3\approx 0$ so that $M_R$ reaches the 
highest order of magnitude possible in these kinds of domains.

Along the regions of Fig.(\ref{fig:cross}) in which $N_2+N_3 \approx l_2+l_3$,
the behavior of $M_R$ depends in a complex way on the difference  
$\vert l_2-l_3 \vert$. While $M_R$ varies
between $4 \ 10^9 {\rm GeV}$ and $2.3 \ 10^{15} {\rm GeV}$ in Fig.(\ref{fig:cross})[a], 
the corresponding values are $2 \ 10^6 {\rm GeV} \lesssim M_R \lesssim 5 \ 10^8 {\rm GeV}$ 
in Fig.(\ref{fig:cross})[d].

In conclusion, as we have shown explicitly for one of them, the two 
solutions of type $N_2,N_3 \approx (l_2+l_3)/2$ and $N_2+N_3 \approx l_2+l_3$, 
leading to a quasi maximum mixing in the neutrino sector (see Section 
\ref{subsec:angles2F}), correspond to maximal $M_R$ values satisfying:  
\begin{equation}
M_R \lesssim 10^{15} {\rm GeV},
\label{eq:limitM2F}
\end{equation} 
for $\Delta m^2_{23}=\Delta m^{2 \ exp}_{23} \approx 3 \ 10^{-3} {\rm eV}^2$ 
\cite{data}. The reason is that for these two kinds of solution, one has 
$\sqrt{V^2_3-V^2_2} \lesssim 1$ so that
$M_R \lesssim \rho^2 / \sqrt{\Delta m^2_{23}} \approx 10^{15} {\rm GeV}$ 
(see Eq.(\ref{DM23bis})) for $\Delta m^2_{23} \approx 3 \ 10^{-3} {\rm eV}^2$.
In other words, Eq.(\ref{DM23}) shows that, when $M_R \approx 10^{15} {\rm GeV}$, 
the needed suppression of $\Delta m^2_{23}=\Delta m^{2 \ exp}_{23}$ compared 
to $\rho^2$ is entirely due to the see-saw mechanism effect (factor 
$\rho^2 / M_R^2$), and, the field localization effect (exponential factor 
in $V^2_3-V^2_2$) does not affect significantly the typical neutrino mass 
scale. Now, for larger values of $M_R$ (too strong mass suppression from 
see-saw mechanism), in order to still have 
$\Delta m^2_{23}=\Delta m^{2 \ exp}_{23}$, one should require that the 
localization effect leads to an increase of $\Delta m^2_{23}$ (see 
Eq.(\ref{DM23})) which is not possible for the two classes of solutions 
considered.

To finish this part, let us remind that the regions of parameter space
$\{e_i,l_j,N_k\}$ presented in Fig.(\ref{fig:cross}) together with the
associated values of $M_R$ given in this section belong to the less 
fine-tuned solutions (described at the end of Section \ref{subsec:angles2F}) 
reproducing all the present experimental data on leptonic sector:  
$m_{\mu^\pm,\tau^\pm}=m^{exp}_{\mu^\pm,\tau^\pm} 
\pm \delta m^{exp}_{\mu^\pm,\tau^\pm}$ \cite{PDG}, 
$0.5<\tan^2\theta_{23}<2.5$ (at $99 \%  \ C.L.$) \cite{data} and
$\Delta m^2_{23}=2.6 \ 10^{-3} {\rm eV}^2$ (best-fit point) \cite{data}.

\subsection{Discussion}
\label{subsec:disres2F}

\subsubsection{Fine-tuning}

In order to discuss quantitatively the fine-tuning of fundamental parameters,
we introduce the following ratio,
\begin{equation}
\bigg \vert {\delta \ln {\cal O} \over \delta \ln {\cal P}} \bigg \vert =
\bigg \vert {\delta {\cal O}/{\cal O} \over \delta {\cal P}/{\cal P}}
\bigg \vert,
\label{eq:FTqty}
\end{equation}
where $\delta {\cal P}$ is the small variation of parameter ${\cal P}$ 
associated to the variation $\delta {\cal O}$ of observable 
${\cal O}$, for any other parameter fixed to a certain value. More physically, 
$\delta {\cal P}$ represents the accuracy required on ${\cal P}$ to 
ensure that ${\cal O}$ is well contained inside its experimental interval 
of width $\delta {\cal O}$.

For instance, let us evaluate the variation ratio defined by 
Eq.(\ref{eq:FTqty}) for the several parameters in the region of parameter 
space given by:
\begin{equation} 
\{e_2=0; e_3\approx 8\mu^{-1}; l_2\approx 4\mu^{-1};
l_3\approx 5\mu^{-1}; N_2\approx 4\mu^{-1}; N_3\in[-15,15]\mu^{-1};
M_R\approx 10^{15}{\rm GeV}\}, 
\label{eq:REGION}
\end{equation}
which is illustrated in Fig.(\ref{fig:cross})[a,b]. 
This domain belongs to the class of solutions (\ref{eq:exsol2F})
fitting the values of $m^{exp}_{e^\pm,\mu^\pm,\tau^\pm}$, and it 
also reproduces the other present experimental results on leptonic sector: 
$0.5<\tan^2\theta_{23}<2.5$ and $\Delta m^2_{23}=2.6 \ 10^{-3} {\rm eV}^2$.
In this domain, the largest quantities of type (\ref{eq:FTqty}) are the 
following ones. From Eq.(\ref{eq:exsol2F}), we obtain an analytical
expression for the partial derivative of $m_{\tau^\pm}$ with respect to 
$e_3$, from which we deduce,
\begin{equation}
\bigg \vert {\delta \ln m_{\tau^\pm} \over \delta \ln e_3} \bigg \vert
=- 2 \ln (m^{exp}_{\tau^\pm}/\rho) \simeq 10.
%=9.89
\label{eq:FTqtyEX2Fa}
\end{equation}
Similarly, Eq.(\ref{eq:exsol2F}) leads to,
\begin{equation}
\bigg \vert {\delta \ln m_{\mu^\pm} \over \delta \ln l_2} \bigg \vert
=- 2 \ln (m^{exp}_{\mu^\pm}/\rho) \simeq 16.
%=15.53
\label{eq:FTqtyEX2Fb}
\end{equation}
From Fig.(\ref{fig:cross})[a,b] and the experimental range: 
$0.5<\tan^2\theta_{23}<2.5$, we derive the two results,
\begin{equation}
\bigg \vert {\delta \ln (\tan^2\theta_{23}) \over \delta \ln l_3} \bigg \vert
\simeq 8,
% 8.04
\
\bigg \vert {\delta \ln (\tan^2\theta_{23}) \over \delta \ln N_2} \bigg \vert
\simeq 4.
% 4.02
\label{eq:FTqtyEX2FdN}
\end{equation}
We also see on Fig.(\ref{fig:cross})[a,b] that $N_3$ can take any value in
the region considered here. Finally, an analytical expression for the 
partial derivative of $\Delta m^2_{23}$ relatively to $M_R$ can be deduced
from Eq.(\ref{DM23}), and the result gives rise to the exact value: 
\begin{equation}
\bigg \vert {\delta \ln (\Delta m^2_{23}) \over \delta \ln M_R} \bigg \vert
=2.
\label{eq:FTqtyEX2Fe}
\end{equation}

\subsubsection{Comparison with the Dirac mass case}

Let us compare these results with the variation ratios of type 
(\ref{eq:FTqty}) obtained in the AS scenario where neutrinos acquire masses 
of Dirac type (SM with an additional right-handed neutrino), which was 
studied in \cite{Bare}. We concentrate on the neutrino sector and consider 
the case where measured leptonic mixing angles originate mainly from this 
sector, as it is what happens both in the present context and in \cite{Bare}.
In the region (\ref{eq:REGION}), all the different ratios (\ref{eq:FTqty}) are 
smaller than the largest ratios (\ref{eq:FTqty}) calculated for any part of the
parameter space associated to AS models where neutrinos have Dirac masses. As 
a matter of fact, within those models, the smallest of four quantities 
$\delta \ln X_i$, where $X_i$ denotes the fundamental parameters: 
$X_i=\{l_2,l_3,N_2,N_3\}$, satisfies $\delta \ln X \lesssim 10^{-2}$ in 
all the different regions of parameter space (see \cite{Bare}). Hence, in the
whole parameter space, the largest ratios (\ref{eq:FTqty}) verify,
\begin{equation}
\bigg \vert {\delta \ln (\tan^2\theta_{23}) \over \delta \ln X} \bigg \vert
\gtrsim 170,
\
\bigg \vert {\delta \ln (\Delta m^2_{23}) \over \delta \ln X} \bigg \vert
\gtrsim 208,
\label{eq:FTqtyDIRbN}
\end{equation}
since the experimental ranges taken in \cite{Bare} are: 
$0.43<\tan^2\theta_{23}<2.33$ and 
$10^{-3} {\rm eV}^2<\Delta m^2_{23}<8 \ 10^{-3} {\rm eV}^2$. The variation ratios of 
Eq.(\ref{eq:FTqtyDIRbN}) can even reach values of the order of 
$1.7 \ 10^6$ and $2.1 \ 10^6$ respectively. The ratios (\ref{eq:FTqtyDIRbN}) 
are much larger than the unity and correspond thus to
an important fine-tuning of parameters, which can even be considered as not 
acceptable \cite{Bare}. Besides, we see that these two ratios are larger than 
the largest ratios (\ref{eq:FTqty}) in region (\ref{eq:REGION}), namely the 
ones of Eq.(\ref{eq:FTqtyEX2FdN}) and Eq.(\ref{eq:FTqtyEX2Fe}).
\\ Within this context, why do we obtain a fine-tuning on $l_2,l_3,N_2$ 
and $N_3$ stronger in the neutrino Dirac mass case (see 
Eq.(\ref{eq:FTqtyDIRbN})) than in the see-saw 
model (see Eq.(\ref{eq:FTqtyEX2FdN})) ?
In the Dirac mass case, the values of $l_2,l_3,N_2$ and $N_3$ are such 
that $\tan^2\theta_{23}$ and $\Delta m^2_{23}$ are in agreement with their
experimental constraints. In contrast, in the see-saw model, $l_{2,3},N_{2,3}$ 
are determined by the experimental interval of $\tan^2\theta_{23}$ (see 
Section \ref{subsec:angles2F}), and then, for each point obtained in
$\{l_2,l_3,N_2,N_3\}$, the other parameter $M_R$ is adjusted in order to 
have a $\Delta m^2_{23}$ value compatible with experimental data (see 
Eq.(\ref{DM23})). In summary, while in the Dirac mass case, two
conditions (reproduction of the experimental values for both 
$\tan^2\theta_{23}$ and $\Delta m^2_{23}$) exist on the parameters 
$l_{2,3},N_{2,3}$, in the see-saw model, there is only one condition (fit of 
the experimental $\tan^2\theta_{23}$ value) that $l_{2,3},N_{2,3}$ must 
fulfill. This difference explains that one finds a fine-tuning on 
$l_{2,3},N_{2,3}$ in the see-saw scenario weaker than in the Dirac mass case.

Note that in the neutrino Dirac mass case, the condition 
$0.43<\tan^2\theta_{23}<2.33$ ($\Leftrightarrow \ \theta_{23} \approx \pi/4$) 
tends to increase greatly the fine-tuning on $l_{2,3},N_{2,3}$. In other 
words, there is a low probability that a neutrino mass matrix (\ref{mnuD}) 
chosen completely at random generates a significant mixing in the neutrino
sector \cite{Bare}. In order to improve the fine-tuning on $l_{2,3},N_{2,3}$,
one could think of reproducing the measured leptonic mixing angle value, 
namely $\theta_{23} \approx \pi/4$, from a large effective mixing in the 
charged lepton sector. Nevertheless, in this case, one would face the same 
problem of strong fine-tuning but now on $l_{2,3},e_3$ (see Eq.(\ref{mlpm})), 
which must also fit the experimental charged lepton masses. 
\\ In the considered see-saw scenario, we have chosen (see 
Sections \ref{subsec:CLmasses2F}, \ref{subsec:angles2F} and 
\ref{subsec:NuMass}) the repartition of experimental constraints on 
fundamental parameters of different sectors
($m_{\mu^\pm,\tau^\pm} \leftrightarrow \{e_i,l_j\}$;
$\tan^2\theta_{23} \leftrightarrow \{l_j,N_k\}$;
$\Delta m^2_{23} \leftrightarrow M_R$)
which allows to improve the situation with regard to the fine-tuning 
status for those parameters.

\subsubsection{Width of $x_5$ range}

We study now the quantity $\Delta X_{MAX}$, namely the largest difference
$\vert X_i-X_j \vert$ where $X_{i,j}=\{l_2,l_3,N_2,N_3\}$. In other terms, 
$\Delta X_{MAX}$ represents the width of $x_5$ range in which all the 
neutrino fields are localized. We compare the $\Delta X_{MAX}$ values, 
that we obtain in the see-saw model, with the $\Delta X_{MAX}$ values found 
in the AS scenario where neutrinos acquire Dirac masses \cite{Bare}. 
As before, and for the same reason, we concentrate on the neutrino sector 
and consider the case where measured leptonic mixing angles originate 
mainly from this sector. For example, in the region 
(Fig.(\ref{fig:cross})[a]), 
\begin{equation} 
\{l_2=3.94\mu^{-1}; l_3=4.5\mu^{-1}; 
N_2\approx N_3\approx 4\mu^{-1};
M_R\approx 10^{15}{\rm GeV}\}, 
\label{eq:REGIONbis}
\end{equation}
which reproduces the present experimental values
$0.5<\tan^2\theta_{23}<2.5$ and $\Delta m^2_{23}=2.6 \ 10^{-3} {\rm eV}^2$
within the see-saw context, we obtain, 
\begin{equation}
\Delta X_{MAX} \approx 0.5\mu^{-1}.
\end{equation}
In contrast, within the case of neutrino Dirac masses, for the 4 possible 
displacement configurations in agreement with $0.43<\tan^2\theta_{23}<2.33$ 
and $10^{-3} {\rm eV}^2<\Delta m^2_{23}<8 \ 10^{-3} {\rm eV}^2$, one has systematically 
\cite{Bare},
\begin{equation}
\Delta X_{MAX} \gtrsim 7.6\mu^{-1}. 
\end{equation}
Notice that the authors of 
\cite{Bare} have imposed a constraint concerning the heaviest neutrino:
$m^2_{\nu_3} \lesssim 10 {\rm eV}^2$, which is also well respected in the final
solutions that we obtain. In conclusion, within the considered see-saw model
and for $M_R$ values much larger than the electroweak scale $\rho$, 
the field positions can be significantly closer to each other than in the 
entire parameter space of AS scenarios where neutrinos acquire Dirac masses. 
The reason was explained in details at the end of Section \ref{subsec:energy}.

\section{Realistic AS Models with 3 Flavors}
\label{sec:3flavors}

\subsection{Charged Lepton Masses}
\label{subsec:CLmasses3F}

\subsubsection{Solutions}

Similarly to the case of 2 lepton flavors,
the localization configurations, consistent with the experimental 
charged lepton masses and corresponding to a minimum fine-tuning of
parameters $l_i$ and $e_j \ [\{i,j\}=1,2,3]$, can be classified into 
144 types of solutions associated to: 6 kinds of mass matrix texture, namely,
\begin{eqnarray}
m^{l^{\pm}}_{I} \simeq
\left ( \begin{array}{ccc}
m_\alpha   & \epsilon_1 & \epsilon_2  \\
\epsilon_3 & m_\beta    & \epsilon_4  \\
\epsilon_5 & \epsilon_6 & m_\gamma
\end{array} \right ), \
m^{l^{\pm}}_{II} \simeq
\left ( \begin{array}{ccc}
\epsilon_1 & m_\alpha   & \epsilon_2  \\
\epsilon_3 & \epsilon_4 & m_\beta     \\
m_\gamma   & \epsilon_5 & \epsilon_6
\end{array} \right ), \
m^{l^{\pm}}_{III} \simeq
\left ( \begin{array}{ccc}
\epsilon_1 & \epsilon_2 & m_\alpha    \\
m_\beta    & \epsilon_3 & \epsilon_4  \\
\epsilon_5 & m_\gamma   & \epsilon_6
\end{array} \right ), 
\cr
m^{l^{\pm}}_{IV} \simeq
\left ( \begin{array}{ccc}
\epsilon_1 & \epsilon_2 & m_\alpha   \\
\epsilon_3 & m_\beta    & \epsilon_4  \\
m_\gamma   & \epsilon_5 & \epsilon_6 
\end{array} \right ), \
m^{l^{\pm}}_{V} \simeq
\left ( \begin{array}{ccc}
\epsilon_1 & m_\alpha   & \epsilon_2  \\
m_\beta    & \epsilon_3 & \epsilon_4  \\
\epsilon_5 & \epsilon_6 & m_\gamma
\end{array} \right ), \
m^{l^{\pm}}_{VI} \simeq
\left ( \begin{array}{ccc}
m_\alpha   & \epsilon_1 & \epsilon_2  \\
\epsilon_3 & \epsilon_4 & m_\beta    \\
\epsilon_5 & m_\gamma   & \epsilon_6 
\end{array} \right ),
\label{finaltextures3F}
\end{eqnarray}
6 mass permutations (distribution of 
$\{m^{exp}_{e^\pm},m^{exp}_{\mu^\pm},m^{exp}_{\tau^\pm}\}$ among 
$[m_\alpha,m_\beta,m_\gamma]$) and 2 opposite signs (for two of 
the differences $l_i-e_j$). These localization configurations 
correspond also to 3 charged lepton mixing 
angles, parameterizing the unitary matrix $U_{lL}$ (defined in 
Eq.(\ref{TRANSmlpm}) and Eq.(\ref{UlL}) for the 2 flavor case),
approximately equal to $0$ or $\pi/2$.

\subsubsection{Symmetries and redundancy}

Any exchange among the fundamental parameters $e_i$, namely 
$e_i \leftrightarrow e_j$, lets all the measurable quantities 
(masses and mixing angles) exactly unchanged. As a matter of fact, it is
equivalent to an exchange between two columns (labeled with the 
indice $j$) of the charged lepton mass matrix $m^{l^{\pm}}_{ij}$ 
(see Eq.(\ref{LAGmlpm}) and Eq.(\ref{mlpm})). Such an exchange 
modifies the matrix $U_{lR}$ (see Eq.(\ref{TRANSmlpm}) for the 2 flavor 
case), on which does not depend the measurable matrix $U_{MNS}$ (see 
Section \ref{subsec:angles2F} for the 2 flavor case), but does not affect 
the hermitian matrix square $m^{l^{\pm}} m^{l^{\pm} \ \dagger}$ and thus 
neither the mixing angles of $U_{lL}$ nor the charged lepton masses 
(see Eq.(\ref{DIAGmlpm}) for the 2 flavor case).

Hence, in our search for realistic localization configurations, 
we do not want to consider solutions identical modulo the symmetry $e_i
\leftrightarrow e_j$, in order to minimize the domains of parameter 
space $\{e_i,l_j,N_k,M_R\}$ which must be studied (nevertheless those
technically identical solutions will all be taken into account in our 
final results). For that purpose, we do not consider the field positions
associated to charged lepton textures $m^{l^{\pm}}_{IV}$, $m^{l^{\pm}}_{V}$ 
and $m^{l^{\pm}}_{VI}$ (see Eq.(\ref{finaltextures3F})) which can be deduced 
respectively from $m^{l^{\pm}}_{II}$, $m^{l^{\pm}}_{III}$ and 
$m^{l^{\pm}}_{I}$ 
via the symmetry $e_2 \leftrightarrow e_3$. The remaining symmetric 
solutions, namely the solutions symmetric under $e_1 \leftrightarrow e_2$ 
and $e_1 \leftrightarrow e_3$, are not considered since we fix the $e_1$ 
value at $e_1=0$, in order to eliminate the physically equivalent solutions 
related to each other by field position translations (see Section 
\ref{subsec:CLmasses2F}). Notice that in the case of 2 lepton flavors (see 
Section \ref{subsec:CLmasses2F}), we have set $e_2$ at zero, for the 
same reason, so that the solutions symmetric under $e_2 \leftrightarrow e_3$
were not considered.

\subsubsection{Explicit examples}

Let us present two explicit examples of localization configurations, 
in agreement
with the observed charged lepton masses, that we find by scanning over 
the parameter space $\{l_1,l_2,l_3,e_2,e_3\}$ ($e_1=0$) with a step of 
$10^{-1}$ for $l_i$ or $10^{-2}$ for $e_2,e_3$ and one of the $l_i$
(the $l_i$ which is approximately
fixed), in the intervals $l_i\in[-15\mu^{-1},15\mu^{-1}]$ and
$e_j\in[-20\mu^{-1},20\mu^{-1}]$ (as suggested by
Eq.(\ref{MassRelationII})).

\begin{figure}[!t]
\begin{center} 
\begin{tabular}{cc}
\psfrag{l3}[c][c][1]{$l_3$}
\psfrag{l2}[c][c][1]{$l_2$}
\includegraphics[width=0.5\textwidth,height=4cm]{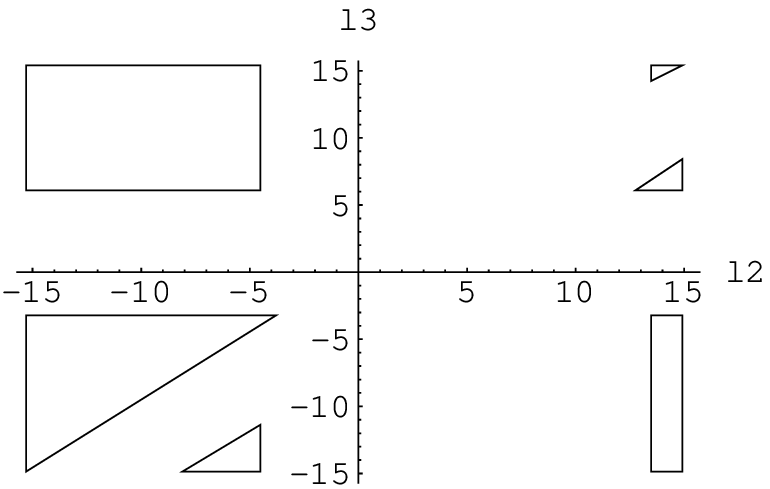} &
\psfrag{l1}[c][c][1]{$l_1$}
\psfrag{l2}[c][c][1]{$l_2$}
\includegraphics[width=0.5\textwidth,height=4cm]{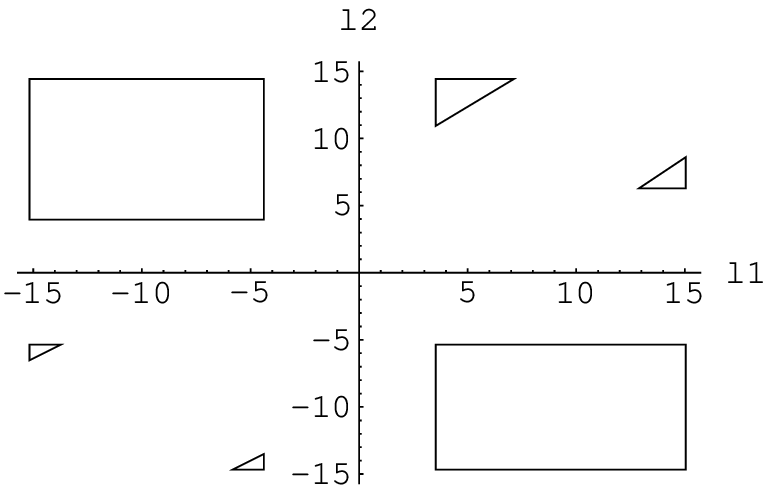} \\
$[a]$ & $[b]$
\end{tabular}
\caption{Domains in the plane $\{l_2,l_3\}$ [a] ($\{l_1,l_2\}$ [b]), 
in units of 
$\mu^{-1}$, reproducing the experimental charged lepton masses, 
together with the values of $e_1,e_2,e_3$ and $l_1$ ($l_3$) given in 
Eq.(\ref{eq:ex1}) (Eq.(\ref{eq:ex2})). Those regions have been derived 
from the scan described in text.}
\label{fig:ex}
\end{center}
\end{figure}

The first example is given by,
\begin{eqnarray}
\bigg \{
e_1=0; \
e_2 \simeq l_2-\mu^{-1} \sqrt{ -2\ln(m^{exp}_{\mu^\pm}/\rho) }; \
e_3 \simeq l_3+\mu^{-1} \sqrt{ -2\ln(m^{exp}_{\tau^\pm}/\rho) }; \cr
l_1 \simeq \mu^{-1} \sqrt{ -2\ln(m^{exp}_{e^\pm}/\rho) }
\bigg \},
\label{eq:ex1}
\end{eqnarray}
with values of $l_2$ and $l_3$ being presented in Fig.(\ref{fig:ex})[a].
This type of solutions is corresponding (see Eq.(\ref{mlpm})) to the texture
$m^{l^{\pm}}_{I}$ of Eq.(\ref{finaltextures3F}) with
$m_\alpha=m^{exp}_{e^\pm}$, $m_\beta=m^{exp}_{\mu^\pm}$,
$m_\gamma=m^{exp}_{\tau^\pm}$ and $e_2-l_2<0$, $e_3-l_3>0$
(the $l_1$ sign is fixed so that the equivalent solutions related by 
the symmetry $X_i\to-X_i$ are eliminated, as explained in Section 
\ref{subsec:CLmasses2F}).

The other example reads as,
\begin{eqnarray}
\bigg \{
e_1=0; \
e_2 \simeq l_1+\mu^{-1} \sqrt{ -2\ln(m^{exp}_{\mu^\pm}/\rho) }; \
e_3 \simeq l_2+\mu^{-1} \sqrt{ -2\ln(m^{exp}_{e^\pm}/\rho) }; \cr
l_3 \simeq \mu^{-1} \sqrt{ -2\ln(m^{exp}_{\tau^\pm}/\rho) }
\bigg \},
\label{eq:ex2}
\end{eqnarray}
with values of $l_1$ and $l_2$ given by Fig.(\ref{fig:ex})[b].
This class of solutions is associated to the texture
$m^{l^{\pm}}_{II}$ of Eq.(\ref{finaltextures3F}) with
$m_\alpha=m^{exp}_{\mu^\pm}$, $m_\beta=m^{exp}_{e^\pm}$,
$m_\gamma=m^{exp}_{\tau^\pm}$ and $e_2-l_1>0$, $e_3-l_2>0$.

In the first example, the values of $l_2$ and $l_3$ can be nearly equal, 
as we remark on Fig.(\ref{fig:ex})[a]. In this situation,
the matrix texture $m^{l^{\pm}}_{I}$ of Eq.(\ref{finaltextures3F}) holds  
typically with (see Eq.(\ref{mlpm})),  
\begin{eqnarray}
\epsilon_4 = m^{l^{\pm}}_{23} \approx m^{l^{\pm}}_{33} \simeq m_\gamma,
\cr
\epsilon_6 = m^{l^{\pm}}_{32} \approx m^{l^{\pm}}_{22} \simeq m_\beta.
\end{eqnarray}
Therefore, the example of Fig.(\ref{fig:ex})[a] shows that the values 
of $\epsilon_i$ parameterizing the textures (\ref{finaltextures3F}), 
which correspond to the less fine-tuned solutions for field displacements, 
can reach orders of magnitude as large as those of experimental 
charged lepton masses ($m_{\alpha,\beta,\gamma}$). In other words,
the domains of parameter space $\{l_1,l_2,l_3\}$, associated to the less
fine-tuned solutions fitting $m^{exp}_{e^\pm,\mu^\pm,\tau^\pm}$, can give 
rise to non-trivial cases where the quantities $\epsilon_i$ of 
Eq.(\ref{finaltextures3F}) are not negligible compared to observed 
charged lepton masses.

\subsection{Mixing Angles and Neutrino Masses}
\label{subsec:angles3F}

\subsubsection{Experimental data}

In this part, we search for the less fine-tuned solutions in parameter
space $\{e_i,l_j,N_k,M_R\}$ which reproduce the experimental values for the
charged lepton masses $m_{e^\pm,\mu^\pm,\tau^\pm}$, the neutrino mass
squared differences $\Delta m^2_{12}=m^2_{\nu_2}-m^2_{\nu_1}$ and
$\Delta m^2_{23}=m^2_{\nu_3}-m^2_{\nu_2}$ (see Eq.(\ref{TRANSmnu}) for the
2 generation case) and the 3 mixing angles $\theta_{12}$, $\theta_{23}$ and 
$\theta_{13}$ parameterizing $U_{MNS}$ (see Eq.(\ref{UmnsI})) as \cite{PDG},
\begin{eqnarray}
U_{MNS}=
\left ( \begin{array}{ccc}
1 &    0    & 0 \\
0 &  c_{23} & s_{23} \\
0 & -s_{23} & c_{23}
\end{array} \right )
\left ( \begin{array}{ccc}
 c_{13} &  0 & s_{13} \\
     0  &  1 & 0 \\
-s_{13} &  0 & c_{13}
\end{array} \right )
\left ( \begin{array}{ccc}
 c_{12} & s_{12} & 0 \\
-s_{12} & c_{12} & 0 \\
   0    &   0    & 1
\end{array} \right ), \cr
U_{MNS}=
\left ( \begin{array}{ccc}
c_{12}c_{13} & s_{12}c_{13} & s_{13} \\
-s_{12}c_{23}-c_{12}s_{23}s_{13}
& c_{12}c_{23}-s_{12}s_{23}s_{13} & s_{23}c_{13} \\
 s_{12}s_{23}-c_{12}c_{23}s_{13}
& -c_{12}s_{23}-s_{12}c_{23}s_{13} & c_{23}c_{13}
\end{array} \right ),
\label{eq:Umns3F}
\end{eqnarray} 
where $s_{ij}=\sin \theta_{ij}$ and $c_{ij}=\cos \theta_{ij}$.
For the present experimental limit on mixing angle $\theta_{13}$, 
which has been obtained from the CHOOZ experiment results \cite{Apollonio:2002gd}, 
we take (see Section \ref{subsec:predict} for the justification of 
this choice),
\begin{equation}
\vert \sin\theta_{13} \vert<0.18.
\label{eq:theta13exp}
\end{equation}
The ranges of values for $\theta_{23}$ and $\Delta m^2_{23}$ consistent  
with the present experimental data, on atmospheric neutrino oscillations and
long-baseline $\nu_\mu$ disappearance from the K2K experiment, are given by 
(at $99 \%  \ C.L.$) \cite{data}\footnote{In \cite{data}, the results on 
neutrino oscillation physics are deduced from a global three lepton 
generation analysis.}, 
\begin{eqnarray}
0.5<\tan^2\theta_{23}<2.5; \ 
\Delta m^2_{23min}<\Delta m^2_{23}<\Delta m^2_{23max}, \cr \cr
\Delta m^2_{23min}=1.5 \ 10^{-3} {\rm eV}^2, \
\Delta m^2_{23max}=3.7 \ 10^{-3} {\rm eV}^2. 
\label{eq:Dmc23exp}
\end{eqnarray}
The present experimentally allowed intervals 
for $\theta_{12}$ and $\Delta m^2_{12}$,
derived from the observations of solar neutrino oscillations and reactor 
$\bar \nu_e$ disappearance with the KamLAND experiment, correspond to the
LMA (large mixing angle) solar neutrino solution \cite{data} 
($99 \%  \ C.L.$):  
\begin{eqnarray}
0.27<\tan^2\theta_{12}<0.9; \
\Delta m^2_{12min}<\Delta m^2_{12}<\Delta m^2_{12max}, \cr \cr
\Delta m^2_{12min}=5 \ 10^{-5} {\rm eV}^2, \
\Delta m^2_{12max}=2 \ 10^{-4} {\rm eV}^2.
\label{eq:Dmc12exp}
\end{eqnarray}

\subsubsection{Method}

{\bf \quad a)} First, in order to find the localization
configurations in agreement with the experimental values for charged lepton 
masses and 3 leptonic mixing angles, 
we have performed a coarse scan over the parameter space 
$\{e_i,l_j,N_k\}$. We have restricted this scan to the regions in 
$\{e_i,l_j\}$ corresponding to the several different types of solutions
fitting experimental charged lepton masses described in Section
\ref{subsec:CLmasses3F}. The ranges considered were 
$l_j\in[-20\mu^{-1},20\mu^{-1}]$, 
$e_i,N_k\in[-25\mu^{-1},25\mu^{-1}]$ (as suggested by 
Eq.(\ref{MassRelationII})), and, the step used was $10^{-1}$ for 
$l_j,N_k$ or $10^{-2}$ for $e_2,e_3$ ($e_1=0$) and the $l_j$ which 
is approximately fixed (see Sections \ref{subsec:CLmasses2F} and
\ref{subsec:CLmasses3F}).

{\bf b)} Then, for each found point of space $\{e_i,l_j,N_k\}$ reproducing 
the correct charged lepton masses and leptonic mixing angles, we have 
determined the values of $M_R$ giving rise to neutrino mass squared
differences contained in their experimental intervals. What are these 
$M_R$ values ? The mass scale $M_R$ satisfies simultaneously the two 
equalities (see Eq.(\ref{DM23bis}) for the 2 family case),
\begin{equation}
M_R= {\rho^2 \sqrt{V^2_2-V^2_1} \over \sqrt{\Delta m^2_{12}}},
\ \mbox{and}, \ 
M_R= {\rho^2 \sqrt{V^2_3-V^2_2} \over \sqrt{\Delta m^2_{23}}},
\end{equation}
where $V_{1,2,3}$ are defined by Eq.(\ref{eq:useful}). 
Hence, in order to have $\Delta m^2_{12}$
and $\Delta m^2_{23}$ values compatible with the experimental results 
(see Eq.(\ref{eq:Dmc23exp}) and Eq.(\ref{eq:Dmc12exp})), $M_R$ must
belong to both the intervals,
\begin{eqnarray}
M^{min}_{R12}<M_R<M^{max}_{R12}, \
M^{min}_{R12}={\rho^2 \sqrt{V^2_2-V^2_1} \over \sqrt{\Delta m^2_{12max}}}, \
M^{max}_{R12}={\rho^2 \sqrt{V^2_2-V^2_1} \over \sqrt{\Delta m^2_{12min}}},
\label{eq:range12}
\end{eqnarray}
and,
\begin{eqnarray}
M^{min}_{R23}<M_R<M^{max}_{R23}, \
M^{min}_{R23}={\rho^2 \sqrt{V^2_3-V^2_2} \over \sqrt{\Delta m^2_{23max}}}, \
M^{max}_{R23}={\rho^2 \sqrt{V^2_3-V^2_2} \over \sqrt{\Delta m^2_{23min}}},
\label{eq:range23}
\end{eqnarray}
which can be summarized as,
\begin{equation}
SUP \bigg ( M^{min}_{R12}, M^{min}_{R23} \bigg ) 
<M_R<
INF \bigg ( M^{max}_{R12}, M^{max}_{R23} \bigg ),
\label{eq:rangesumup}
\end{equation}
$SUP$ ($INF$) standing for superior (inferior).
Some of the points found in $\{e_i,l_j,N_k\}$, fitting the correct 
values for $m_{e^\pm,\mu^\pm,\tau^\pm}$, $\theta_{12}$, $\theta_{23}$ and
$\theta_{13}$, lead to distinct intervals 
(\ref{eq:range12}) and (\ref{eq:range23}) without a common range (namely such 
that $SUP ( M^{min}_{R12}, M^{min}_{R23} )
>INF ( M^{max}_{R12}, M^{max}_{R23} )$). For those points, there
exist no intervals of $M_R$ (as in Eq.(\ref{eq:rangesumup})) consistent with 
the experimental limits on $\Delta m^2_{12}$ and $\Delta m^2_{23}$. 
Therefore, these points are not selected for the final solutions in 
$\{e_i,l_j,N_k,M_R\}$ reproducing the experimental values of charged 
lepton masses, leptonic mixing angles and neutrino mass squared 
differences (see Eq.(\ref{eq:theta13exp}), Eq.(\ref{eq:Dmc23exp}) and 
Eq.(\ref{eq:Dmc12exp})). In contrast, notice that in the 2 lepton family 
case, the experimental constraint on neutrino mass squared difference 
did not bring any additional test, or condition, on field positions.

\subsubsection{Results}

By the method described just above, we find that the less fine-tuned 
solutions in parameter space $\{e_i,l_j,N_k,M_R\}$ reproducing all the
present experimental data on lepton sector ($m_{e^\pm,\mu^\pm,\tau^\pm}
= m^{exp}_{e^\pm,\mu^\pm,\tau^\pm}$ plus the intervals of 
Eq.(\ref{eq:theta13exp}), Eq.(\ref{eq:Dmc23exp}) and
Eq.(\ref{eq:Dmc12exp})) consist of 10 distinct and limited domains,
denoted as $A_i,B_i,C_i,D_i,E_i \ [i=1,2]$. Those domains are described
in details in Appendix \ref{FINALsol}.

\begin{figure}[!t]
\begin{center}

\begin{tabular}{c}

\includegraphics[width=\textwidth]{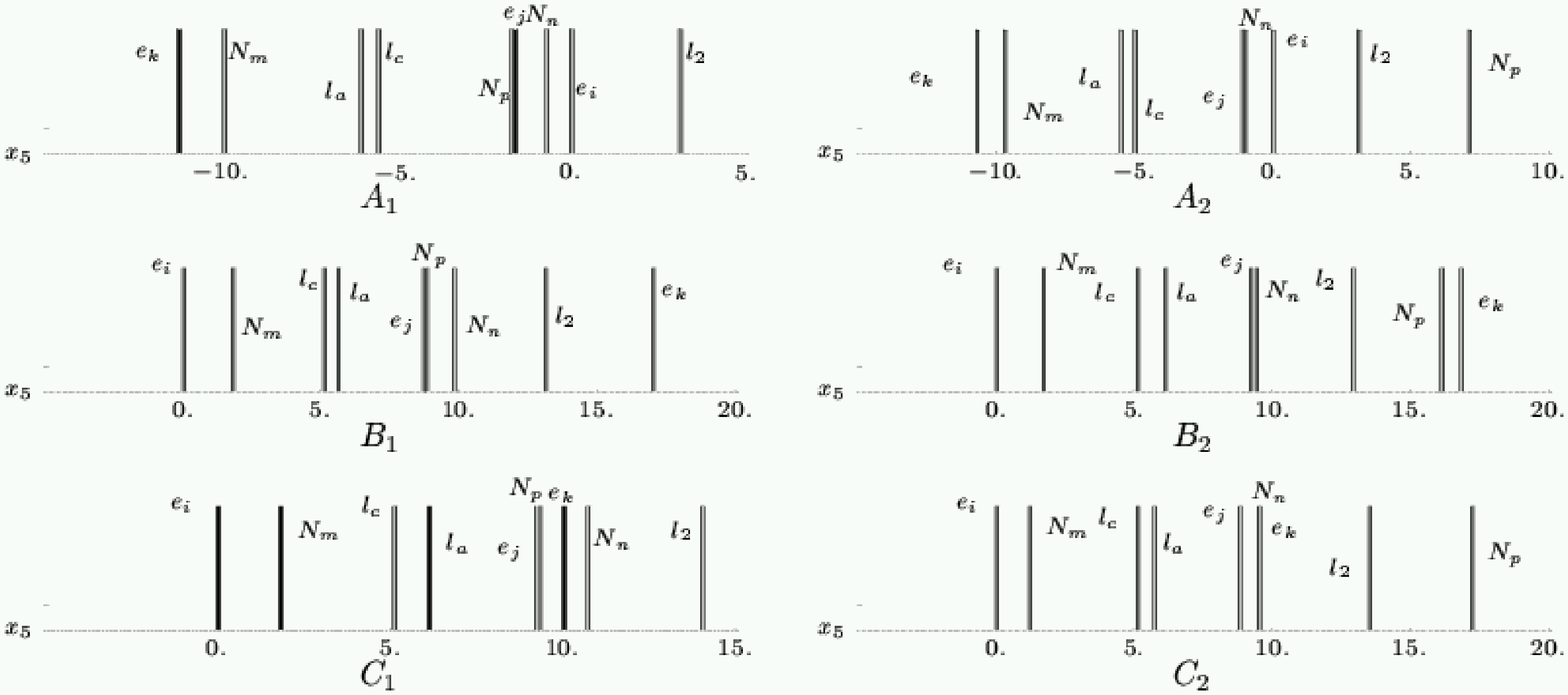}\\
\includegraphics[width=\textwidth]{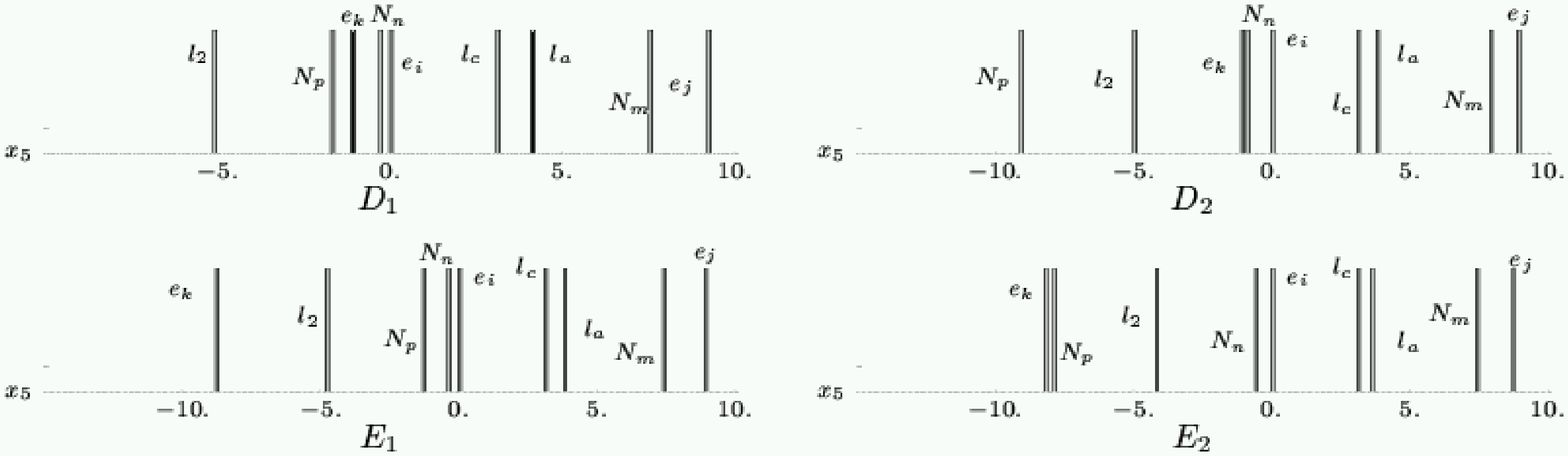}
\end{tabular}

\caption{Complete field position configurations along the extra dimension
(parameterized by $x_5$ in units of $\mu^{-1}$) associated to one point
in each of the domains $A_{1,2},B_{1,2},C_{1,2},D_{1,2},E_{1,2}$ (see
text) of parameter space $\{e,l,N,M_R\}$. The $e$ indices $\{i,j,k\}$
(distributed among $[1,2,3]$) and $l$ indices $\{a,c\}$ (distributed
among $[1,3]$) obey to the same indice notation as the one adopted in
Eq.(\ref{eq:10regEL}). Thus, some of the distances, between $l$ and $e$
positions, shown here are given by Eq.(\ref{eq:10regEL}). Similarly, the
$N$ indices $\{m,n,p\}$ (distributed among $[1,2,3]$) correspond to the
indices of Eq.(\ref{eq:5regLN}), Eq.(\ref{eq:5regLNBIS}) and
Eq.(\ref{eq:pattern}), so that the configurations of $l$ and $N$
presented here are quantified in those relations. Each point has been
chosen in order to have $l_2$ given by the value associated to highest
peak of its distribution (for the relevant domain).}
\label{fig:heavy}
\end{center}
\end{figure}

\begin{figure}[!t] 
\begin{center}
\begin{tabular}{cc}
\psfrag{l1}[c][c][1]{$l_{1}$}
\psfrag{l2}[c][c][1]{$l_{2}$}
\includegraphics[height=4cm,width=0.5\textwidth]{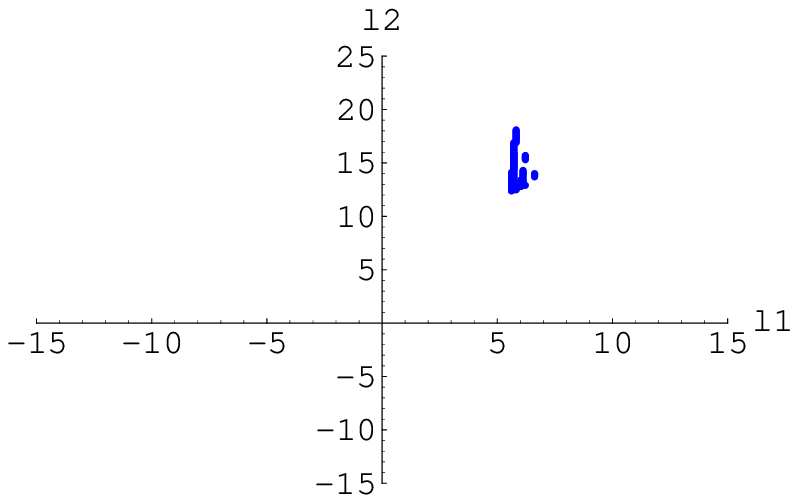}&
\psfrag{l1}[c][c][1]{$l_{1}$}
\psfrag{N1}[c][c][1]{$N_{1}$}
\includegraphics[height=4cm,width=0.5\textwidth]{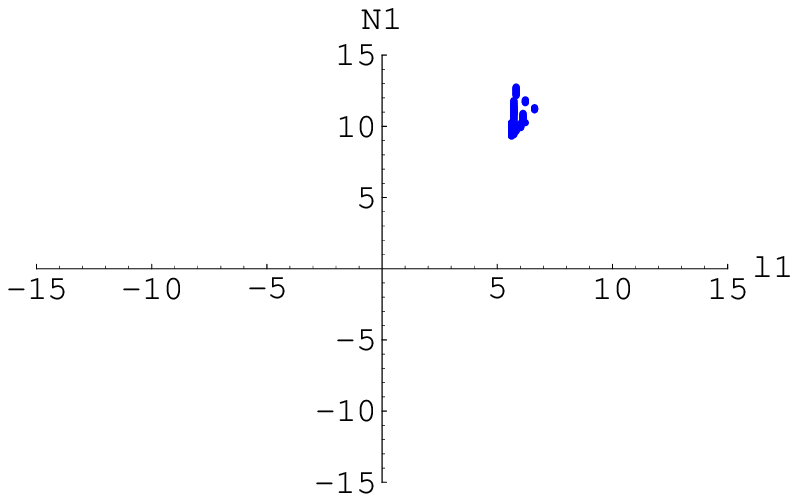}\\
 & \\
\psfrag{N1}[c][c][1]{$N_{1}$}
\psfrag{N2}[c][c][1]{$N_{2}$}
\includegraphics[height=4cm,width=0.5\textwidth]{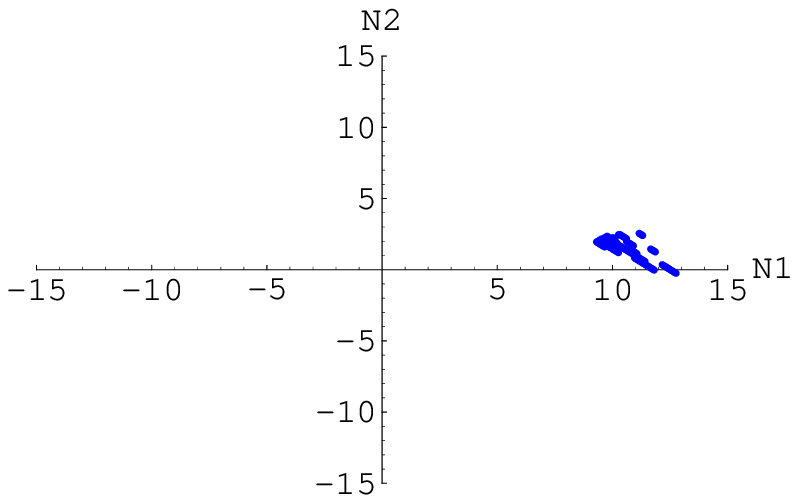}&
\psfrag{N1}[c][c][1]{$N_{1}$}
\psfrag{N3}[c][c][1]{$N_{3}$}
\includegraphics[height=4cm,width=0.5\textwidth]{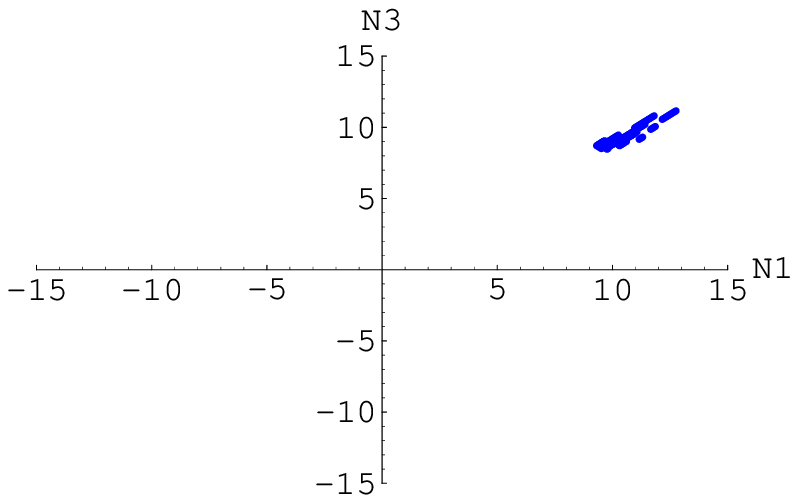}
\end{tabular}
\caption{Fundamental parameters $l_{1,2}$ and $N_{1,2,3}$ 
(units of $\mu^{-1}$) in the domain $B_1$ (see text) for which 
$0=e_1 \simeq l_3-\mu^{-1} \sqrt{ -2\ln(m^{exp}_{e^\pm}/\rho) }$, 
$e_2 \simeq l_1+\mu^{-1} \sqrt{ -2\ln(m^{exp}_{\tau^\pm}/\rho) }$, 
$e_3 \simeq l_2+\mu^{-1} \sqrt{ -2\ln(m^{exp}_{\mu^\pm}/\rho) }$ 
(we take $i=1$, $j=2$, $k=3$, $a=1$, $c=3$ 
in Eq.(\ref{eq:10regEL})), and, $N_2 \approx {3l_b-l_2 \over 2}$, 
$N_1 \approx N_3 \approx {l_b+l_2 \over 2}$ (we choose $m=2$, 
$n=1$, $p=3$ in Eq.(\ref{eq:5regLN})).} 
\label{fig:complete}
\end{center}
\end{figure}

In order to illustrate these results, we show explicitly, 
on Fig.(\ref{fig:heavy}), a given point in each of the 10 distinct 
regions $A_i,B_i,C_i,D_i,E_i \ [i=1,2]$ of parameter space 
$\{e_i,l_j,N_k,M_R\}$. Furthermore, on Fig.(\ref{fig:complete}), 
we present the entire region in $\{e_i,l_j,N_k\}$ associated to $B_1$.

Based on the physical interpretation of the 2 flavor case given in 
Section \ref{subsec:angles2F}, we discuss now the various mixing angles
associated to the obtained solutions $A_i,B_i,C_i,D_i,E_i \ [i=1,2]$. 
\\ Let us begin with an explanation of the two types of characteristic
configurations (\ref{eq:5regLN}) and (\ref{eq:5regLNBIS}) for 
positions $l_i,N_j$. For simplification reasons,
we take for instance $m=1$, $n=2$ and $p=3$ in both Eq.(\ref{eq:5regLN}) 
and Eq.(\ref{eq:5regLNBIS}). Then, Eq.(\ref{eq:5regLNBIS}) can be 
rewritten as,
\begin{eqnarray}
N_1 \approx {3l_1-l_2 \over 2}, \
N_2 \approx {l_1+l_2 \over 2}, 
\label{eq:5regLNBISfixI}
\end{eqnarray}
\begin{eqnarray}
N_2 \approx {l_2+l_3 \over 2}, \
N_3 \approx {3l_2-l_3 \over 2}.
\label{eq:5regLNBISfixII}
\end{eqnarray}
These typical configurations (\ref{eq:5regLNBISfixI}) and 
(\ref{eq:5regLNBISfixII}) are equivalent to the configuration 
(\ref{eq:SOLcross2F}), with respectively $N_1$ and $N_3$ far from $N_2$ 
(see Fig.(\ref{fig:cross})[c,d]), which leads to a nearly maximum mixing 
in the neutrino sector within the context of 2 flavors. Hence, the 
configurations (\ref{eq:5regLNBISfixI}) and (\ref{eq:5regLNBISfixII}) give
rise to the neutrino mixing angle values: $\theta_{12}^\nu \approx \pi/4$ 
and $\theta_{23}^\nu \approx \pi/4$. Indeed, the sectors $\nu_e-\nu_\mu$ 
and $\nu_\mu-\nu_\tau$ decouple from each other, due to a small value for
$\theta_{13}^\nu$, and can thus be treated as separated 2 flavor sectors. 
More precisely, in Eq.(\ref{eq:5regLNBISfixI}) and 
Eq.(\ref{eq:5regLNBISfixII}), 
while the relation involving $N_2$ leads to $\theta_{12}^\nu \approx \pi/4$
and $\theta_{23}^\nu \approx \pi/4$ (like in Eq.(\ref{eq:SOLcross2F})), the 
relations on $N_1$ and $N_3$ give rise to $\theta_{13}^\nu \approx 0$ as well 
as field positions compatible with the experimental neutrino mass squared 
differences.
\\ Similarly, for $m=1$, $n=2$ and $p=3$, Eq.(\ref{eq:5regLN}) can be
rewritten:
\begin{eqnarray}
N_1 \approx {3l_1-l_2 \over 2}, \
N_2 \approx {l_1+l_2 \over 2},
\label{eq:5regLNfixI}
\end{eqnarray}
\begin{eqnarray}
N_2 \approx {l_2+l_3 \over 2}, \
N_3 \approx {l_2+l_3 \over 2}.
\label{eq:5regLNfixII}
\end{eqnarray}
The typical configuration (\ref{eq:5regLNfixII}) is identical to 
$N_2+N_3 \approx l_2+l_3$ (see Eq.(\ref{eq:SOLdiag2F})), for 
$N_2 \approx N_3$ (see Fig.(\ref{fig:cross})[c,d]), leading to a quasi 
maximal mixing in the neutrino sector if only 2 flavors are considered. 
As before, in Eq.(\ref{eq:5regLNfixI}) and Eq.(\ref{eq:5regLNfixII}),
the relations involving $N_2$ and $N_3$ lead thus to 
$\theta_{12}^\nu \approx \pi/4$ and $\theta_{23}^\nu \approx \pi/4$, 
whereas the formula in which enters $N_1$ gives rise to 
$\theta_{13}^\nu \approx 0$ and allows the neutrino mass squared differences 
to have correct values.
\\ To sum up, the characteristic 3 flavor configuration of $l_i,N_j$ 
given by Eq.(\ref{eq:5regLN}) is based on the two typical configurations
(\ref{eq:SOLcross2F}) and (\ref{eq:SOLdiag2F}) described in details in 
Section \ref{subsec:angles2F} and leading both to a quasi 
maximal mixing for the neutrino sector within a 2 flavor framework. 
In contrast, the typical 3 flavor configuration (\ref{eq:5regLNBIS}) 
is only based on the 2 flavor configuration (\ref{eq:SOLcross2F})
explained and interpreted in Section \ref{subsec:angles2F}.
\\ We conclude from this discussion that the parameters in regions
$A_i,B_i,C_i,D_i,E_i \ [i=1,2]$, which obey to the relations 
(\ref{eq:5regLN}) if $i=1$ and (\ref{eq:5regLNBIS}) if $i=2$, lead to 
$\theta_{12}^\nu \approx \pi/4$, $\theta_{23}^\nu \approx \pi/4$
and $\theta_{13}^\nu \approx 0$. Now, the solutions 
$A_i,B_i,C_i,D_i,E_i \ [i=1,2]$ reproduce the wanted values for the 
3 mixing angles parameterizing $U_{MNS}$ (see Eq.(\ref{eq:Umns3F})), namely
$\vert \theta_{12} \vert \approx \pi/4$ (Eq.(\ref{eq:Dmc12exp})), 
$\vert \theta_{23} \vert \approx \pi/4$ (Eq.(\ref{eq:Dmc23exp})) and 
$\theta_{13} \approx 0$ (Eq.(\ref{eq:theta13exp})). Therefore (see
Eq.(\ref{UmnsI})), in the domains $A_i,B_i,C_i,D_i,E_i \ [i=1,2]$, the 3 
mixing angles of charged lepton sector, which parameterize $U_{lL}$, are 
approximately equal to either $0$ or $\pi/2$. It means that there is an
almost vanishing effective mixing in the charged lepton sector for the 
typical localization configurations of $e_i,l_j$ (consistent with 
$m_{e^\pm,\mu^\pm,\tau^\pm} = m^{exp}_{e^\pm,\mu^\pm,\tau^\pm}$) given 
by Eq.(\ref{eq:10regEL}) (classes of solutions described in 
Section \ref{subsec:CLmasses3F}), as we said in 
Section \ref{subsec:CLmasses3F}.

\subsubsection{Conclusion}

In summary of this section, we have found the less fine-tuned solutions 
in parameter space $\{e_i,l_j,N_k$, $M_R\}$ in agreement 
with all the present experimental 
data on charged lepton masses, neutrino mass squared differences and  
leptonic mixing angles (Eq.(\ref{eq:theta13exp}), Eq.(\ref{eq:Dmc23exp}) and
Eq.(\ref{eq:Dmc12exp})). These solutions can be classified into some of the 
different types of $e_i,l_j$ configurations defined in 
Section \ref{subsec:CLmasses3F}, which correspond to a nearly vanishing
effective mixing in the charged lepton sector. In fact, these solutions
obey to the characteristic relations (\ref{eq:5regLN}) or (\ref{eq:5regLNBIS}) 
on field positions $l_j,N_k$, leading both to neutrino mixing angles 
responsible for almost the whole leptonic mixing observed experimentally.
We have explained and interpreted those final solutions, and we have 
presented (description in Appendix \ref{FINALsol} together with 
Fig.(\ref{fig:heavy})) all 
the associated regions of parameter space $\{e_i,l_j,N_k$,$M_R\}$ 
($A_{1,2},B_{1,2},C_{1,2},D_{1,2},E_{1,2}$), except the ranges of values for 
the typical Majorana mass scale $M_R$ that we will give and discuss in 
Section \ref{subsec:predict}.

\subsection{Discussion}
\label{subsec:disc3F}

\subsubsection{Fine-tuning}

We estimate here the variation ratio of type (\ref{eq:FTqty}) for the 
different parameters in domain $B_1$ (see Section \ref{subsec:angles3F}), 
for example. Remind that the solution $B_1$ reproduces all the present
experimental data on leptonic sector: the experimental values for 
$m^{exp}_{e^\pm,\mu^\pm,\tau^\pm}$ and the constraints of 
Eq.(\ref{eq:theta13exp}), Eq.(\ref{eq:Dmc23exp}) and Eq.(\ref{eq:Dmc12exp}).
As in Fig.(\ref{fig:complete}), we consider the domain $B_1$ characterized by 
$i=1$, $j=2$, $k=3$, $a=1$, $c=3$ (see Eq.(\ref{eq:10regEL})) and 
$m=2$, $n=1$, $p=3$ (see Eq.(\ref{eq:5regLN})). The largest quantities 
(\ref{eq:FTqty}) are the following ones. From Eq.(\ref{eq:10regEL}), 
we obtain an analytical expression for the partial derivative of 
$m_{\tau^\pm}$ relatively to $e_2$, from which we deduce,
\begin{equation} 
\bigg \vert {\delta \ln m_{\tau^\pm} \over \delta \ln e_2} \bigg \vert
=- 2 \ln (m^{exp}_{\tau^\pm}/\rho) \simeq 10.
%=9.89
\label{eq:FTqtyEX3Fa}
\end{equation}
Similarly, Eq.(\ref{eq:10regEL}) leads to,
\begin{equation}
\bigg \vert {\delta \ln m_{\mu^\pm} \over \delta \ln e_3} \bigg \vert
=- 2 \ln (m^{exp}_{\mu^\pm}/\rho) \simeq 16,
%=15.53
\ 
\bigg \vert {\delta \ln m_{e^\pm} \over \delta \ln l_3} \bigg \vert
=- 2 \ln (m^{exp}_{e^\pm}/\rho) \simeq 26.
%=26.25
\label{eq:FTqtyEX3FcN}
\end{equation}
Within the region $B_1$, and for the smallest $N_1$ values, the quantity 
$\delta \ln X_i$, where $X_i=\{l_1,l_2,N_1,N_2,N_3\}$, is about 
$\delta \ln X_i \sim 10^{-1}$, as illustrated by Fig.(\ref{fig:complete}). 
Hence, based on Eq.(\ref{eq:theta13exp}), Eq.(\ref{eq:Dmc23exp}) and 
Eq.(\ref{eq:Dmc12exp}), we find, 
\begin{equation}
\bigg \vert {\delta \ln (\sin \theta_{13}) \over \delta \ln X_i} \bigg \vert
\simeq 20,
\
\bigg \vert {\delta \ln (\tan^2\theta_{23}) \over \delta \ln X_i} \bigg \vert
\simeq 16,
\
\bigg \vert {\delta \ln (\tan^2\theta_{12}) \over \delta \ln X_i} \bigg \vert
\simeq 12,
\label{eq:FTqtyEX3FfN}
\end{equation}
\begin{equation}
\bigg \vert {\delta \ln (\Delta m^2_{23}) \over \delta \ln X_i} \bigg \vert
\simeq 9,
\
\bigg \vert {\delta \ln (\Delta m^2_{12}) \over \delta \ln X_i} \bigg \vert
\simeq 14.
\label{eq:FTqtyEX3FhN}
\end{equation}
Finally, in the part of $B_1$ where the $N_1$ values are minimum, one has 
$10^{4}{\rm GeV} \lesssim M_R \lesssim 10^{12}{\rm GeV}$ and the variation of $M_R$
(entire width of interval (\ref{eq:rangesumup})), associated to the variations 
of $\Delta m^2_{23}$ and $\Delta m^2_{12}$ within their respective 
experimental range, reads as $\delta \ln M_R \simeq 0.43$. We thus find 
(see Eq.(\ref{eq:Dmc23exp}) and Eq.(\ref{eq:Dmc12exp})),
\begin{equation}
\bigg \vert {\delta \ln (\Delta m^2_{23}) \over \delta \ln M_R} \bigg \vert
\simeq 2,
%2.10 
\
\bigg \vert {\delta \ln (\Delta m^2_{12}) \over \delta \ln M_R} \bigg \vert
\simeq 3. 
%3.22
\label{eq:FTqtyEX3FjN}
\end{equation}

\subsubsection{Perturbative bound}

Let us make here a fundamental comment concerning the quantity 
$\Delta X_{MAX}$ (introduced in Section \ref{subsec:disres2F}), 
namely the largest difference $\vert X_i-X_j \vert$ where 
$X_{i,j}=\{e_{1,2,3},l_{1,2,3},N_{1,2,3}\}$. Inside the whole   
domains of parameter space obtained 
($A_{1,2},B_{1,2},C_{1,2},D_{1,2},E_{1,2}$), which fit all the present 
experimental data on leptonic sector, one has systematically,  
\begin{equation}
15\mu^{-1} \lesssim \Delta X_{MAX} \lesssim 30\mu^{-1},
\label{eq:DXconclu}
\end{equation} 
as can be deduced from 
Eq.(\ref{eq:10regEL}), Eq.(\ref{eq:5regLN}), Eq.(\ref{eq:5regLNBIS}),
Eq.(\ref{eq:rangeli}) and Eq.(\ref{eq:rangeliBIS}). This result on 
$\Delta X_{MAX}$ can be clearly seen in Fig.(\ref{fig:complete}), for 
the example of entire region $B_1$. By consequence (since 
$\Delta X_{MAX} \lesssim L$), for all our final solutions in parameter space 
$\{e_{1,2,3},l_{1,2,3},N_{1,2,3},M_R\}$, the important constraint on wall 
thickness $L \lesssim 30\mu^{-1}$ (see Eq.(\ref{MassRelationII})), coming 
from considerations concerning perturbativity, can be well respected.

\vskip 0.5cm

We end up this part by noting that the less fine-tuned consistent 
realizations of the AS scenario, that we obtain (regions 
$A_{1,2},B_{1,2},C_{1,2},D_{1,2},E_{1,2}$ of parameter space),  
do not recover the particular field localization configuration 
proposed in \cite{Raidal:2002xf} for justifying a minimal see-saw model.

\section{Predictions}
\label{subsec:predict}

In this part, we will describe some features and predictions for neutrino 
sector, within the see-saw model, which are provided by the complete and 
consistent realizations of AS scenario obtained in 
Section \ref{subsec:angles3F}. We will have a look on
the light left-handed neutrino sector and then
on the heavy right-handed one. Our solutions are quite
predictive on the mixing angle $\theta_{13}$ and lead to a 
more precise prediction on the 
lightest neutrino mass eigenvalue $m_{\nu_1}$. So we will
compare our predictions on those two physical quantities with the
corresponding present and future neutrino experiment sensitivities.
We will finish by important comments on the value of parameter $\rho$.

We will first consider the case of normal hierarchy for light left-handed 
neutrino masses (as before), and the inverted hierarchy case will be discussed 
after in a separate section.

\subsection{Normal Mass Hierarchy}
\label{subsec:rho}

\begin{figure}[!t]
\begin{center} 
\begin{tabular}{cc}
\psfrag{m1}[c][c][1]{{\myblue $m_{\nu_1}$}}
\psfrag{m2}[c][c][1]{{\mygreen $m_{\nu_2}$}}
\psfrag{m3}[c][c][1]{{\myred $m_{\nu_3}$}}
\psfrag{lmi}[c][r][1][90]{$Log_{10}[m_{\nu_i}\ {\rm (eV)}]$}
\psfrag{logMGeV}[c][r][1]{$Log_{10}[M_R\ {\rm ({\rm GeV})}]$}
\includegraphics[width=0.5\textwidth,height=5cm]{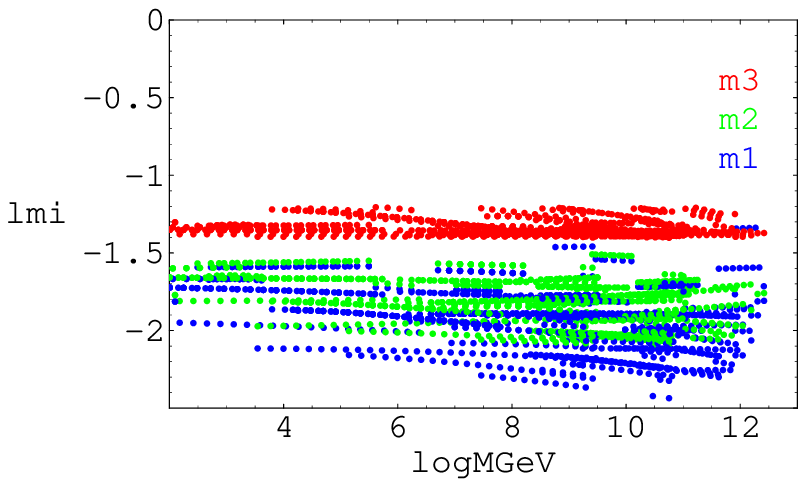}&
\psfrag{myF1}[c][c][1][90]{$m_{\nu_1}\ ({\rm eV})$}
\psfrag{m1max}[c][c][1]{{\scriptsize {\myblue $m^{min}_{\nu_1}$}}}
\psfrag{m1min}[c][c][1]{{\scriptsize {\mygreen $m^{max}_{\nu_1}$}}}
\psfrag{LogMGeV}[c][r][1]{$Log_{10}[M_R\ {\rm (GeV)}]$}
\includegraphics[width=0.5\textwidth,height=5cm]{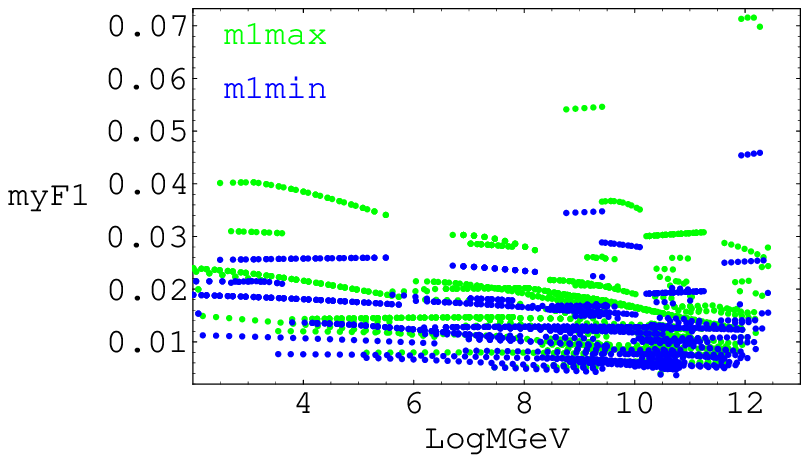}\\

[a]&[b]

\end{tabular}
 \caption{\small  Light neutrino masses $m_{\nu_1}$, $m_{\nu_2}$ and $m_{\nu_3}$ (eV) [a] and 
 boundaries of the  allowed range for $m_{\nu_1}$ (eV) [b] as a
 function of right-handed Majorana neutrino mass $M_R$ (GeV). The $m_{\nu_i}$
 values in [a] correspond to the lower limit of allowed mass 
 range (in particular, $m_{\nu_1}$ in [a] corresponds to $m_{\nu_1}^{min}$ in [b]). 
 The values shown in [a] and [b] correspond to
 the solutions $A,B,C,D,E|_{1,2}$ defined in Appendix \ref{FINALsol}.}
\label{fig:mi} 
\end{center}
\end{figure}

\subsubsection{Light left-handed sector}

The $M_R$ values being given by the range (\ref{eq:rangesumup}),  
we derive from Eq.(\ref{eq:useful}) the ground of light neutrino masses $m_{\nu_1}$, {\it i.e.}
the corresponding range for the lightest neutrino: $m_{\nu_1}\in [m^{min}_{\nu_1},m_{\nu_1}^{max}]$, where,
\begin{equation}
m_{\nu_1}^{max}=INF\left( \sqrt{\Delta m^2_{12max}}
{V_1 \over \sqrt{V^2_2-V_1^2}}\ ,\ \sqrt{\Delta m^2_{23max}}
{V_1 \over \sqrt{V^2_3-V_2^2}}   \right),
\label{m1max}
\end{equation}
and,
\begin{equation}
m_{\nu_1}^{min}=SUP\left(  \sqrt{\Delta m^2_{12min}}
{V_1 \over \sqrt{V^2_2-V_1^2}}\ ,\ \sqrt{\Delta m^2_{23min}}  
{V_1 \over \sqrt{V^2_3-V_2^2}}   \right).
\label{m1min}
\end{equation}
Our solutions exhibit a quite remarkable feature. Indeed, for all our final
models, though running together over several orders of magnitude, $V_1$, $V_2$
and $V_3$ remain actually of the same order and such that the ratios
$V_1/\sqrt{V^2_2-V_1^2}$ and $V_1/\sqrt{V^2_3-V_2^2}$ are close to
one, leading to nearly constant bounds $m^{min}_{\nu_1}$ and
$m^{max}_{\nu_1}$. Furthermore, those two bounds are systematically close, 
which gives rise to the generic prediction: 
$$
m_{\nu_1}\sim10^{-2} {\rm eV}, 
$$ 
and consequently to the typical spectrum:
$$
 m_{\nu_1} \sim m_{\nu_2}\lesssim m_{\nu_3}.
$$
This picture is illustrated on Fig.(\ref{fig:mi}). A more physical way to
understand the constance of $m_{\nu_1}$ in our solutions is the following: 
there is a compensation in Eq.(\ref{eq:useful}) between the variations of 
$M_R$ (entering the suppression factor $\rho/M_R$ due to the see-saw
mechanism) and $V_i$ (depending on the suppression factors issued from
wave function overlaps).

Besides, our solutions correspond to $0.05 {\rm eV}< \sum
m_{i}<0.12 {\rm eV}$ which agrees with the upper cosmological bound
$\sum m_{i} <0.7-1.01 {\rm eV}$ (depending on cosmological priors)
coming from WMAP and 2dFGRS galaxy survey \cite{Hannestad:2003xv}.

\subsubsection{Heavy right-handed sector}

\begin{figure}[!t]
\begin{center} 
\begin{tabular}{cc}
\psfrag{M1}[c][c][1]{{\myblue $M_{1}$}}
\psfrag{M2}[c][c][1]{{\mygreen $M_{2}$}}
\psfrag{M3}[c][c][1]{{\myred $M_{3}$}}
\psfrag{lMi}[c][r][1][90]{$Log_{10}[M_i\ {\rm (GeV)}]$}
\psfrag{logMGeV}[c][r][1]{$Log_{10}[M_R\ {\rm (GeV)}]$}
\includegraphics[width=0.5\textwidth,height=5cm]{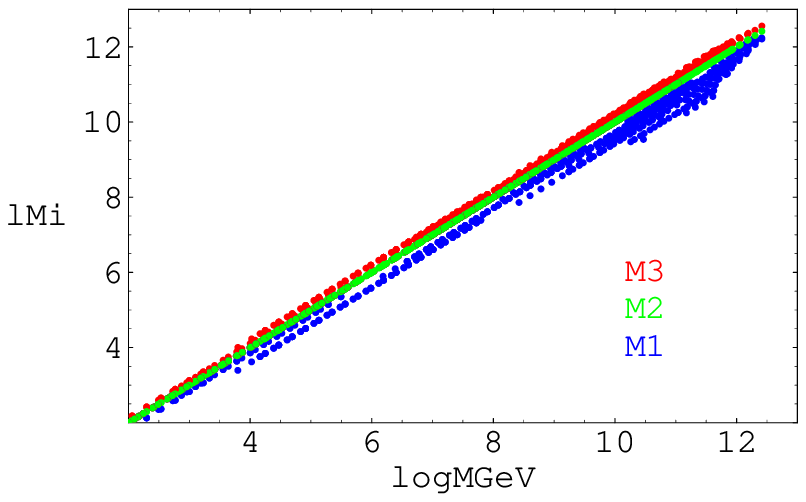}&
\psfrag{ABCDE1}[c][r][1][1]{\scriptsize {{\myblue $A,B,C,D,E|_1$}}}
\psfrag{ABCDE2}[c][r][1][1]{\scriptsize {{\myred $A,B,C,D,E|_2$}}}
\psfrag{l1N3}[c][r][1][90]{$\vert l_1-N_3 \vert \ (\mu^{-1})$}
\psfrag{logMGeV}[c][r][1]{$Log_{10}[M_R\ {\rm (GeV)}]$}
\includegraphics[width=0.5\textwidth,height=5cm]{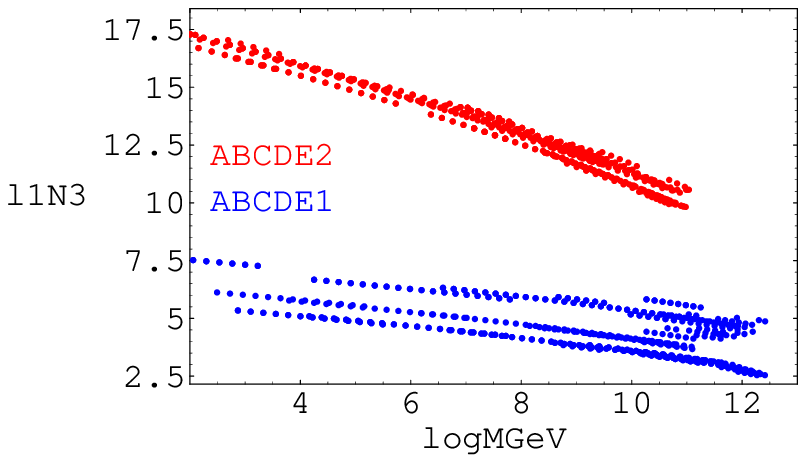}\\

[a]&[b]

\end{tabular}
 \caption{\small [a] Right-handed neutrino masses $M_1,M_2,M_3$ (GeV) as
 function of the common scale $M_R$ (GeV) (Eq.(\ref{M})). 
 [b] Correlation of  $M_R$ with the field position difference
 $\vert l_1-N_3 \vert$ (in units of $\mu^{-1}$). The values shown
 in [a] and [b] correspond to the solutions $A,B,C,D,E|_{1,2}$ defined in 
 Appendix \ref{FINALsol}.}
\label{fig:Mi} 
\end{center}
\end{figure}

Lepton masses and mixings fix the positions
$\{e_i,l_j,N_k\}$ accordingly to configurations such that $N_1,N_2$ and $N_3$ 
are quite far from each other (see Fig.(\ref{fig:heavy})). This leads
to a quasi diagonal Majorana mass matrix $M_{ij}$ (Eq.(\ref{M})), up to
essentially one non-diagonal term $M_{np}$ ($n$ and
$p$ correspond to the $N$ indices in Fig.(\ref{fig:heavy})) which can reach
$\sim 0.5 M_R$ in solutions $A,B,C,D,E|_1$. This texture gives rise to 
three eigenvalues $M_i$, for the Majorana mass matrix 
$M_{ij}$ (Eq.(\ref{M})), of the same order of magnitude: 
$$
M_1\sim M_2 \sim M_3 \sim M_R,
$$
especially in the second type of solutions 
$A,B,C,D,E|_2$ where, $\vert N_n-N_p \vert$ being higher (see
Fig.\ref{fig:heavy}), one has an almost
exact degeneracy. This is illustrated on Fig.(\ref{fig:Mi})[a].

Let us now discuss the $M_R$ values associated to all our solutions.
We have chosen to present in Fig.(\ref{fig:mi}) and Fig.(\ref{fig:Mi})
the $M_R$ values corresponding to the upper
boundary of allowed range (\ref{eq:rangesumup}), the two
boundaries being actually close ($0.65<{\rm ratio}<1$)
for each point of our solutions. When field positions vary, due to the 
presence of exponential factors, the eigenvalues
$V_i$ vary strongly so that $M_R$ spans several orders of
magnitude (see Eq.(\ref{eq:rangesumup})). 
As we want to stay in the see-saw mechanism spirit, we
impose as a minimum $M_R\gtrsim 10^2{\rm GeV}$, which explains the
lower $M_R$ axis boundary on Fig.(\ref{fig:mi}) and Fig.(\ref{fig:Mi}). 
\\ In contrast, the upper bound appearing on Fig.(\ref{fig:mi}) 
and Fig.(\ref{fig:Mi}), namely:
$$
M_R \lesssim 10^{12}{\rm GeV},
$$ 
is a consequence of the 
localization configurations we have obtained. Let us try to understand 
this bound. Due to the small value of mixing angle $\theta_{13}$, 
the sectors $\nu_e-\nu_\mu$ and $\nu_\mu-\nu_\tau$ 
decouple. Hence, we deduce from the discussion within the 2 lepton
flavor case (end of Section \ref{subsec:NuMass}) that the quantities
$M^{max}_{R12}$ (defined in Eq.(\ref{eq:range12})) and $M^{max}_{R23}$
(defined in Eq.(\ref{eq:range23})) satisfy to  
$M^{max}_{R12} \lesssim 10^{16} {\rm GeV}$ and 
$M^{max}_{R23} \lesssim 10^{15} {\rm GeV}$ (see Eq.(\ref{eq:limitM2F})).
Nevertheless, $M^{max}_{R12}$ and $M^{max}_{R23}$ do not reach their
upper value in the same region of parameter space $\{e_i,l_j,N_k\}$ (otherwise
Eq.(\ref{eq:rangesumup}) would lead to $M_R < INF(M^{max}_{R12},M^{max}_{R23})
\lesssim INF(10^{16} {\rm GeV},10^{15} {\rm GeV}) = 10^{15} {\rm GeV}$) which explains why
the resulting upper limit on $M_R$, namely $\sim 10^{12}{\rm GeV}$,
is smaller than $\sim 10^{15}{\rm GeV}$.
\\ Why the upper limit on $M_R$ is higher for solutions $A,B,C,D,E|_1$ than
$A,B,C,D,E|_2$ ? In solutions of type 1, the position difference
$\vert l_b-N_p \vert$ [$b=1,3$] is systematically smaller than in configurations of type 2,
as it is clear on Fig.(\ref{fig:heavy}). Therefore, the eigenvalues
$V_i$ are typically larger in solutions
of type 1 than in solutions of type 2. The consequence 
(see Eq.(\ref{eq:rangesumup})) is that
$M_R$ can reach larger values in solutions of type 1. This argument is
illustrated in Fig.(3)[b] for the case: $b=1$ and $p=3$.

Finally, our result of three right-handed Majorana masses remaining 
of the same order, over ten orders of magnitude, is interesting
with regard to the leptogenesis. We postpone leptogenesis, in this 
AS framework based on the see-saw mechanism, for a next coming paper 
\cite{ournext}.

\subsubsection{Comparison with experimental sensitivities}

Today, there are still two measurable quantities (forgetting relevant phases) 
undetermined (actually weakly constrained), in the
light neutrino sector: the ground of neutrino masses (namely $m_{\nu_1}$, in the
normal hierarchy case with our conventions) and the
mixing angle $\theta_{13}$. For those two physical quantities, we are going to
briefly remind the experimental status and then compare the experimental
sensitivities with our predictions.

Concerning the lightest neutrino mass, the present
measurements of kinematics in tritium $\beta$ decay 
give rise to the bound $m_{\nu_1}<2.2{\rm eV}$ (Mainz \cite{Lobashev:tp} and
Troisk \cite{Bonn:tw} Collaborations). This result will be improved by
the KATRIN project \cite{Osipowicz:2001sq} down to $m_{\nu_1}<0.35{\rm
eV}$.
\\ Furthermore, neutrinoless double $\beta$ decay (see
\cite{Elliott:2002xe} for a recent review) experiments
are sensitive to the so-called effective neutrino mass: $m_{ee}=|\sum_i
U^2_{ei}m_{\nu_i}|$ ($U_{ei}$ are the first line elements of $U_{MNS}$ 
matrix (\ref{eq:Umns3F})).
Those measurements are complementary with those on $m_{\nu_1}$
because they require neutrinos to be Majorana particles. 
The Heidelberg-Moscow 
experiment \cite{Heidel} obtains the lowest limit: $m_{ee}<0.35{\rm
eV}$. This bound will be improved by next generation experiments like
EXO \cite{EXO}, MOON \cite{MOON}, XMASS \cite{XMASS},
NEMO3 \cite{Sarazin:2000xv}, CUORE \cite{Arnaboldi:2002du} and GENIUS
\cite{Klapdor-Kleingrothaus:2000ue}, down to $m_{ee}\lesssim 10^{-2}{\rm
eV}$ (highest expected sensitivity) for the latter.

Concerning the $\theta_{13}$ mixing angle, $s_{13}$ (same notation as
in Eq.(\ref{eq:Umns3F})) is up to now constrained by the CHOOZ experiment
\cite{Apollonio:2002gd}. For the $\Delta m^2_{13}$ value that we obtain: 
$\sqrt{\Delta m^2_{13}} \sim 5 \ 10^{-2}{\rm eV}$ (see Fig.(\ref{fig:mi})), the 
bound from CHOOZ experiment is approximately $s_{13}<0.18$ \cite{Apollonio:2002gd}, 
as we have imposed in our analysis\footnote{Note that the bound 
$s_{13} \lesssim 0.18$ corresponds also to the limit obtained at $90 \%  \ C.L.$
in the global three neutrino flavor analysis of \cite{data}.} 
(see Eq.(\ref{eq:theta13exp})).
This bound on $s_{13}$ will be improved (see {\it e.g.}
\cite{Huber:2002mx,Huber:2003pm}) by superbeams like CNGS
\cite{Acquistapace:1998rv}, JHF \cite{Itow:2001ee} and neutrino
factories \cite{NuFact}. Depending on possible improvement, 
JHF-SK, JHF-HK (or low luminosity 
neutrino factories which have quite similar performances on $s_{13}$ 
measurement) and finally high luminosity neutrino factories will at
least improve the bound down to respectively \cite{Huber:2002mx}: 
$s_{13} \lesssim 7 \  10^{-2}$, $1.6 \  10^{-2}$, $1.1 \  
10^{-2}$.

\begin{figure}[!t]
\begin{center} 
\begin{tabular}{cc}
\psfrag{se3}[c][r][1]{$s_{13}$}
\psfrag{logm1eV}[c][r][1]{$Log_{10}[m_{\nu_1}\ {\rm (eV)}]$}
\psfrag{katrin}[c][c][1]{{\scriptsize KATRIN}}
\psfrag{mainz}[c][c][1]{{\scriptsize Mainz,}}
\psfrag{troisk}[c][ct][1]{{\scriptsize Troisk}}
\psfrag{JHF}[l][l][1]{{\scriptsize JHF-SK}}
\psfrag{nuF}[l][l][1]{{\scriptsize $\nu$F2}}
\includegraphics[width=0.5\textwidth,height=5cm]{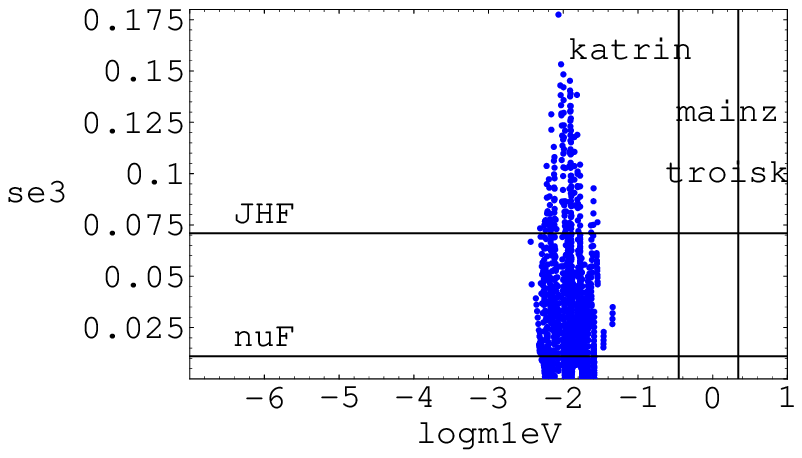}&
\psfrag{C}[l][l][1]{{\scriptsize CHOOZ}}
\psfrag{J}[l][l][1]{{\scriptsize JHF-SK}}
\psfrag{J2}[l][l][1]{{\scriptsize JHF-HK/$\nu$F1}}
\psfrag{nuF}[l][l][1]{{\scriptsize $\nu$F2}}
\psfrag{H}[l][l][1]{{\scriptsize Heidel}}
\psfrag{N3}[l][l][1]{{\scriptsize NEMO3}}
\psfrag{G}[l][l][1]{{\scriptsize GENIUS}}
\psfrag{lse3}[c][r][1][90]{$Log_{10}[s_{13}]$}
\psfrag{logmeeeV}[c][r][1]{$Log_{10}[m_{ee}\ {\rm (eV)}]$}
\psfrag{-2.1}[c][r][1]{}
\psfrag{-1.9}[c][r][1]{}
\psfrag{-1.7}[c][r][1]{}
\psfrag{-1.5}[c][r][1]{}
\includegraphics[width=0.5\textwidth,height=5cm]{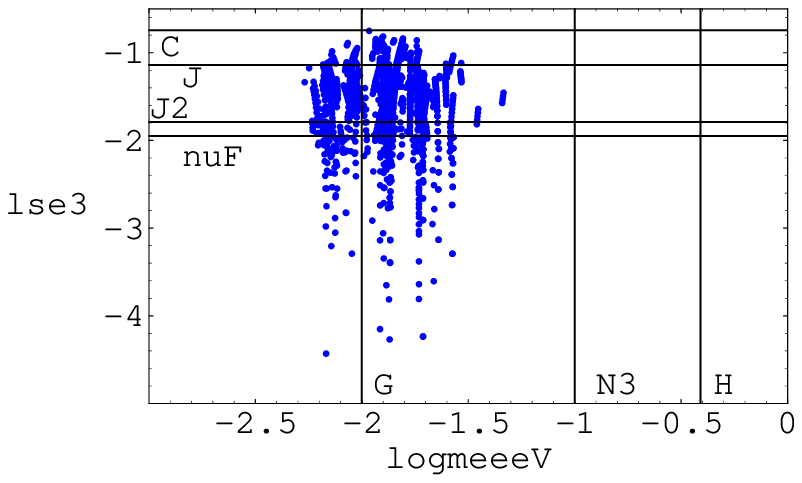}\\

[a]&[b]

\end{tabular}
 \caption{\small [a] $(m_{\nu_1},s_{13})$ plane with present 
 (Mainz and Troisk) and future (KATRIN) experimental sensitivities on $m_{\nu_1}$. [b]
 $(m_{ee},s_{13})$ plane with current (Heidelberg) and future 
 (NEMO3, GENIUS) bounds on $m_{ee}$ from double $\beta$ decay experiments. 
 Current (CHOOZ) and future
 superbeams (JHF-SK, JHF-HK) and neutrino factories (low luminosity:
 $\nu$F1, high luminosity: $\nu$F2) sensitivities on $s_{13}$ are also
 represented. The values shown in [a,b] correspond to the solutions 
 $A,B,C,D,E|_{1,2}$ and to the lower limit of neutrino mass 
 interval (for instance, the $m_{\nu_1}$ value is taken equal to $m_{\nu_1}^{min}$).}
\label{fig:m1se3} 
\end{center}
\end{figure}

All those sensitivities are indicated on Fig.(\ref{fig:m1se3}) where we
also show all our solutions: $A,B,C,$ $D,E|_{1,2}$ in the planes $(s_{13},m_{\nu_1})$ 
and $(s_{13},m_{ee})$. Our models are not extremely predictive on
$s_{13}$ although they ``prefer'' the range: 
$$
10^{-2} \lesssim s_{13} \lesssim 10^{-1},
$$ 
to which belongs most of our final solutions.  
It is interesting to remark that this range 
contains the best-fit value $s_{13}\simeq 0.1$ obtained 
in the global three neutrino oscillation analysis of \cite{data}.
Furthermore, this favored range will be directly tested by JHF-SK(HK) and
neutrino factories. Concerning the
masses, the relatively precise prediction on $m_{\nu_1}$ and $m_{ee}$, in
the AS scenario based on the see-saw mechanism, is particularly clear
on Fig.(\ref{fig:m1se3}). The
predicted value of $m_{\nu_1}$ is out of reach of Mainz and Troisk current
bound and will not be reached by the future Katrin sensitivity
(Fig.(\ref{fig:m1se3})[a]). Nevertheless, assuming zero-phases in $m_{ee}$ 
(our $U_{ei}$ being real), our solutions
lead to the value $m_{ee}\sim 10^{-2}{\rm eV}$ which will be
accessible by some future double $\beta$ decay experiments. 
Indeed, the GENIUS experiment should
be able to test a wide set of our solutions, as shown in 
Fig.(\ref{fig:m1se3})[b]. Hence, future measurements
on $\theta_{13}$ and $m_{\nu_1}$ will allow to either motivate or 
exclude most of our theoretical models.

\subsubsection{Choice of the $\rho$ value}

As explained in Section \ref{subsec:CLmasses2F}, within the considered 
AS scenario, the parameter $\rho=\kappa(m_{top})\langle h \rangle$ (defined in 
Eq.(\ref{mlpm})) must be of the order of magnitude of the electroweak scale,
since one should have $\kappa(m_{top}) \approx 1$. Hence, one could 
assume an exact $\rho$ value a bit different from the one we have chosen, 
namely $\rho=1.5 m_{top}(m_{top})$ (see Eq.(\ref{rhomtop})), as long as 
$\rho$ remains of the same order of magnitude of the electroweak scale. The 
important message that we deliver here is that such a different choice of 
$\rho$ would not have significantly affected our predictions on neutrino 
sector (both in the normal and inverted mass hierarchy cases).

As it is clear from Eq.(\ref{eq:rangesumup}), a modification of $\rho$ leads 
to a variation of $M_R$. Nevertheless, for a small modification of $\rho$
(such that $\rho$ remains of the order of electroweak scale), the $M_R$
values do not vary significantly.

Besides, within the present framework, the variations of left-handed neutrino
masses $m_{\nu_i}=m_{\nu_i}(l_j,N_k)$ (given by Eq.(\ref{m1max}) and Eq.(\ref{m1min})), 
due to a modification of $\rho$, would only originate from the variations 
of $V_i(l_j,N_k)$ eigenvalues (defined in Eq.(\ref{eq:useful})). Indeed, the 
obtained values of parameter $l_j$ (and $e_i$), and thus of $V_i(l_j,N_k)$, 
depend on $\rho$ (see Eq.(\ref{mlpm})).
\\ We find numerically that for a small modification of $\rho$ ($\rho$ staying 
around the order of magnitude of electroweak scale), the variations of all 
neutrino masses $m_{\nu_i}(l_j,N_k)$ are negligible. It is due to the fact that the 
constraints on parameters $l_j,N_k$ (and $e_i$), issued from requiring lepton 
mixing angles compatible with their experimental bounds, are exactly invariant 
under a modification of $\rho$\footnote{The lepton mixing angles do not depend 
on any overall factor of lepton mass matrices (as can be checked for instance 
in Eq.(\ref{eq:tg2tetanu})), so that in the considered framework, they do 
not depend directly on $\rho$ (see Eq.(\ref{mlpm}) and 
Eq.(\ref{SeesawFormulaAS})).}.

\subsection{Inverted Mass Hierarchy}

Here, we discuss the inverted hierarchy case which is characterized by: 
$$ 
m_{\nu_3}<m_{\nu_2}<m_{\nu_1}, 
$$ 
with $\Delta m^2_{23}=m_{\nu_2}^2-m_{\nu_3}^2$
given by atmospheric neutrino oscillation data, and,
$\Delta m^2_{12}=m_{\nu_1}^2-m_{\nu_2}^2$
given by solar neutrino oscillation data.
For this neutrino mass hierarchy, our models lead to the same picture
for $s_{13}$ and again three left-handed neutrino masses quite close
to each other. 
However, this time the
prediction is on $m_{\nu_3}$ (with the above notation):
$$
m_{\nu_3}\sim10^{-2}{\rm eV}, 
$$ 
with the following spectrum:
$$ \ m_{\nu_3} \lesssim m_{\nu_1} \sim m_{\nu_2} .$$
In the right-handed sector, the picture of three heavy masses of
the same order of magnitude, namely 
$M_i \sim M_R$, is unchanged. Furthermore, due to the  suppression of
$m_{\nu_3}$ by $U_{e3}^2$ (small $s_{13}$ value) in the $m_{ee}$ expression, 
$m_{ee}$ is a bit higher in the inverted hierarchy case than in  the
normal hierarchy case. The cloud of our solutions would thus be centered on 
$m_{ee}\sim 4 \  10^{-2}{\rm eV}$ in the inverted hierarchy case 
(instead of $m_{ee}\sim 10^{-2}{\rm eV}$ as in
the normal hierarchy case of Fig.(\ref{fig:m1se3})[b]), which implies
that the GENIUS experiment would be able to test almost all our solutions.

\section{Conclusion}

We have studied the structure of lepton flavor space in the framework of the AS
scenario with an extra dimension, under the assumption that neutrinos acquire 
Majorana masses via the see-saw mechanism. \\ First, we have given a generic 
description of the AS models reproducing the correct charged lepton masses 
and corresponding to a minimal fine-tuning of parameters. 
Then, we have shown that one can construct realistic AS models, namely 
compatible with all the present experimental data on the entire leptonic 
sector (charged lepton masses, and, constraints on mixing angles and neutrino 
masses), in which neutrinos acquire masses through the see-saw 
mechanism. We have found the realistic AS models of this kind giving rise to
a minimum fine-tuning of the fundamental parameters (lepton field positions 
and neutrino Majorana mass scale $M_R$). Those AS models can be classified 
into some of the different types of localization configurations defined in 
Section \ref{subsec:CLmasses3F}, and, verify either the characteristic 
relation (\ref{eq:5regLN}) or (\ref{eq:5regLNBIS}). Furthermore, all these
AS models, which have been explained and interpreted, lead to neutrino mixing 
angles responsible for almost the whole effective leptonic mixing measured 
experimentally.

We have found that, within the considered AS framework based on the see-saw 
mechanism, the minimal fine-tuning on parameters (reached when examining all  
the relevant parameter space) 
is weaker than in the AS scenario where neutrinos acquire Dirac masses 
\cite{Bare}. This is due to the fact that the parameters of our model are
those of the AS scenario with neutrino Dirac masses (lepton field positions) 
plus an additional parameter ($M_R$). In the AS scenario with neutrino Dirac 
masses, there is an important fine-tuning of parameters, which can be 
considered as not acceptable \cite{Bare}. In contrast, we have shown that 
some of our final solutions associated to the see-saw model (as for instance 
$B_1$), fitting all the experimental results on charged leptons and neutrinos, 
correspond to an acceptable fine-tuning of parameters. It means that, if 
nature has chosen the AS scenario in order to generate the whole lepton flavor 
structure, the case in which neutrino masses are produced via the 
see-saw mechanism seems to be the most favored one, as it can lead to a 
reasonable fine-tuning of parameters.

Moreover, we have demonstrated that, in the AS scenarios based on
see-saw model that we have obtained, for $M_R$ values much greater than 
the electroweak symmetry breaking scale, the lepton field positions can be 
significantly closer to each other than in the whole parameter space of AS 
scenarios with neutrino Dirac masses \cite{Bare}. In other terms, the 
see-saw mechanism is in favor of the naturality of the AS scenario, in 
the sense given in Section \ref{intro}. Besides, we have obtained the 
important result that, even when considering the entire leptonic sector, 
the constraint on domain wall thickness along the extra dimension: 
$L \lesssim 30\mu^{-1}$, coming from considerations concerning perturbativity, 
can be well respected in concrete realizations of the AS scenario based on 
the see-saw model (as it occurs in all our final solutions).

One of our major quantitative results is the following one. The less fine-tuned 
realistic AS models based on the see-saw mechanism, that we have obtained, give 
rise to predictions on the whole neutrino sector. Indeed, our final models lead 
to three right-handed neutrino Majorana masses of the same order of magnitude: 
$$
M_1 \sim M_2 \sim M_3 \sim M_R,
$$ 
and which can run between $\sim 10^2{\rm GeV}$ and $\sim 10^{12}{\rm GeV}$.
Moreover, those models predict a ground level for left-handed neutrino 
Majorana masses, leading to the following typical mass spectrum,
\begin{eqnarray}
m_{\nu_1} \sim m_{\nu_2} \sim 10^{-2}{\rm eV} < m_{\nu_3} \sim 5 \ 10^{-2}{\rm eV}
& \quad {\rm [normal \ hierarchy]},\nonumber 
\\
m_{\nu_3} \sim 10^{-2}{\rm eV} <  m_{\nu_2} \sim m_{\nu_1} \sim 5 \ 10^{-2}{\rm eV}
& \quad {\rm [inverted \ hierarchy]}.\nonumber
\end{eqnarray}
Most of the values found for the effective neutrino mass $m_{ee}$,
which correspond to this predicted spectrum, should be testable in  
future double $\beta$ decay experiments. Finally, our AS models favor
the range of $\theta_{13}$ mixing angle value defined by,
$$
10^{-2} \lesssim s_{13} \lesssim 10^{-1},
$$
which should be accessible by future superbeam and neutrino factory 
sensitivities.

In fact, the concrete realization of AS scenario based on the see-saw 
mechanism that we have elaborated, together with the AS model in \cite{Mira}
which fits all quark masses and mixing angles, constitute the first 
realistic and complete realization of the AS scenario with one extra 
dimension. 
A complete realization in the sense that it reproduces all the present
experimental data on SM fermions, and a realistic one in the three following
senses. First, this realistic realization allows a reasonable fine-tuning
of parameters. Secondly, in this realization, the experimental bounds on 
proton lifetime and FCNC rates are well respected. These bounds are
satisfied thanks to the existence of large mass scales (see 
Section \ref{subsec:energy}), which is possible as we have assumed that 
the fundamental energy scale is high ($M_\star^{5D} \gtrsim 10^{16} {\rm GeV}$). 
Thirdly, in this physical realization, the width of $x_5$ interval 
along the extra dimension, inside which all lepton (quark) fields are 
localized, ranges approximately between $15\mu^{-1}$ ($20\mu^{-1}$) and 
$30\mu^{-1}$ ($25\mu^{-1}$), as shown in our relation (\ref{eq:DXconclu}) 
(in \cite{Mira}). Therefore, if both the lepton and quark fields are trapped  
in the same region of extra dimension, then the perturbativity condition on 
domain wall thickness $L \lesssim 30\mu^{-1}$ can be fulfilled.

How the hypothesis of a fundamental scale smaller than the GUT scale 
would have modify our study ? For $M_\star^{5D} \lesssim 10^{16} {\rm GeV}$,
the interactions mediating proton decay (and FCNC reactions) should be 
partially suppressed through a localization of quarks and leptons (and 
SM fermions of different families) at far positions from each other, the 
suppression due to large mass scales being not sufficient (see 
Section \ref{subsec:energy}). For instance, if $M_\star^{5D} \sim 1 {\rm TeV}$, 
the proton can be stable enough, for a distance between quark and lepton
positions in the extra dimension of at least $\sim 10\mu^{-1}$ \cite{AS}. 
Those lower bounds on distances between field positions make that 
the perturbativity condition on domain wall width $L \lesssim 30\mu^{-1}$ 
is not necessarily (or easily) satisfied when one considers all the 
SM fermions, depending on the $M_\star^{5D}$ value and localization 
configuration. In particular, the FCNC constraints on field positions 
could even be not compatible with the experimental data on SM fermions.

Finally, let us comment on the other hierarchy problem in the SM: the 
instability of the Higgs boson mass, and thus the electroweak energy scale, 
with respect to quantum corrections. 
In a reasonable realization of the AS 
scenario with only one extra dimension, this hierarchy problem does not
seem to be able to be solved by bringing the Planck scale down to just above 
the electroweak scale (as proposed in \cite{ADD}). Indeed, as discussed
previously, in complete realistic AS models with one extra dimension, the 
width of $x_5$ range, in which all fields are localized, must be larger than 
at least $\sim 10\mu^{-1}$ typically (for instance, in the quark model of
\cite{Mira}, it is larger than $\sim 20\mu^{-1}$), so that one has 
$L \gtrsim 10\mu^{-1}$. Now, this bound on $L$ together with the constraint
characteristic of the AS scenario, $\mu^2 L < M_\star^{5D}$ (see 
Section \ref{subsec:energy}), imply the typical limit on fundamental 
scale: $M_\star^{5D} \gtrsim 100 L^{-1} \gtrsim 100 {\rm TeV}$, as $L^{-1}$ 
cannot be pushed significantly below roughly $1 {\rm TeV}$. Nevertheless, one 
could try to find a particular realization of AS scenario with more than 
one extra dimension, which would allow to greatly reduce the $M_\star^{5D}$ 
value. 
Anyway, if the hierarchy problem of Higgs mass instability is solved by 
bringing the Planck scale down to $M_\star^{5D}={\cal O}(1){\rm TeV}$, then the 
motivation for the see-saw model becomes weaker. As a matter fact, in such 
a situation, the Majorana mass scale is bounded by $M_R \lesssim 1{\rm TeV}$, so 
that the see-saw mechanism cannot be responsible for a significant 
suppression of left-handed neutrino masses relatively to the electroweak scale 
($\rho$: see Eq.(\ref{SeesawFormulaAS})) and looses thus its important 
r\^ole with regard to neutrino mass generation.  
Therefore, in realizations of the AS scenario, based on the see-saw 
mechanism, it is desirable that the hierarchy problem can be solved by a
model compatible with a $M_\star^{5D}$ value larger than the electroweak 
scale. For example, the Higgs mass can be stabilized in a supersymmetric
theory softly broken at the electroweak scale, because of the cancellation
of all quadratically divergent quantum corrections.

\vspace{1cm}

\noindent {\bf \Large Acknowledgments}

\noindent The authors are grateful to P.~Aliani, M.~V.~Libanov and 
J.~Orloff for interesting discussions. G.~M. and E.~N. acknowledge 
support from the Belgian Federal Government under contract IAP V/27, 
the French Community of Belgium (ARC) and the IISN.

\newpage

\appendix
\noindent {\bf \Large Appendix}
\vspace{0.5cm}

\renewcommand{\thesubsection}{A.\arabic{subsection}}
\renewcommand{\theequation}{A.\arabic{equation}}
\setcounter{subsection}{0}
\setcounter{equation}{0}

\section{Neutrino Mass Matrix Expression}
\label{ExMatrix2F}

In the case of two lepton flavors ($\{i,j\}=2,3$), the expression for 
left-handed neutrino mass matrix generated by the see-saw mechanism 
within an AS context (namely the Majorana mass matrix of 
Eq.(\ref{SeesawFormulaAS})) reads as,   
\begin{displaymath}
m^{\nu}_{ij} \simeq 
\left ( \begin{array}{cc}
m^{\nu}_{22} & m^{\nu}_{23} \\ 
m^{\nu}_{23} & m^{\nu}_{33}
\end{array} \right ), \ \mbox{with}, \ 
\end{displaymath} 
\begin{eqnarray}
m^{\nu}_{22} & = & 
- {\rho^2 \over M_R (1-e^{-\mu^2(N_2-N_3)^2})} \bigg (
e^{-\mu^2(l_2-N_2)^2}+e^{-\mu^2(l_2-N_3)^2} \cr &&
-2e^{-{\mu^2\over 2}[(N_2-N_3)^2+(l_2-N_2)^2+(l_2-N_3)^2]}
\bigg ), \cr
m^{\nu}_{33} & = &
- {\rho^2 \over M_R (1-e^{-\mu^2(N_2-N_3)^2})} \bigg (
e^{-\mu^2(l_3-N_2)^2}+e^{-\mu^2(l_3-N_3)^2} \cr && 
-2e^{-{\mu^2\over 2}[(N_2-N_3)^2+(l_3-N_2)^2+(l_3-N_3)^2]} 
\bigg ), \cr
m^{\nu}_{23} & = &
- {\rho^2 \over M_R (1-e^{-\mu^2(N_2-N_3)^2})} \bigg (
e^{-{\mu^2\over 2}[(l_3-N_2)^2+(l_2-N_2)^2]}
+e^{-{\mu^2\over 2}[(l_3-N_3)^2+(l_2-N_3)^2]}
\cr &&
-e^{-{\mu^2\over 2}[(N_2-N_3)^2+(l_3-N_2)^2+(l_2-N_3)^2]}
-e^{-{\mu^2\over 2}[(N_2-N_3)^2+(l_2-N_2)^2+(l_3-N_3)^2]}
\bigg ).\cr && 
\label{eq:ExMa2F}
\end{eqnarray}

\renewcommand{\thesubsection}{B.\arabic{subsection}}
\renewcommand{\theequation}{B.\arabic{equation}}
\setcounter{subsection}{0}
\setcounter{equation}{0}

\section{Final 3 Flavor Solutions}
\label{FINALsol}

The less fine-tuned solutions in parameter space $\{e_i,l_j,N_k,M_R\}$ 
fitting all the present experimental data on leptons ($m_{e^\pm,\mu^\pm,\tau^\pm}
= m^{exp}_{e^\pm,\mu^\pm,\tau^\pm}$ together with the ranges of 
Eq.(\ref{eq:theta13exp}), Eq.(\ref{eq:Dmc23exp}) and
Eq.(\ref{eq:Dmc12exp})) constitute ten distinct and limited regions,
that we denote as $A_i,B_i,C_i,D_i,E_i \ [i=1,2]$. In this appendix, 
we describe those regions in details.

First, the regions $A_i,B_i,C_i,D_i,E_i \ [i=1,2]$ 
correspond to some of the different types of localization configurations 
consistent with the experimental charged lepton masses described in 
Section \ref{subsec:CLmasses3F}. Indeed, the points in these domains have
coordinates obeying to the following respective relations ($\{i,j,k\}$ are 
distributed among $[1,2,3]$, and, $\{a,c\}$ among $[1,3]$),
\begin{eqnarray}
A_{1,2}: 
\ \ \ \ \ \ \ \ \ \ \ \ \ \ \ \ \ \ \ \ \ \ \ \ \ \ \ 
\ \ \ \ \ \ \ \ \ \ \ \ \ \ \ \ \ \ \ \ \ \ \ \ \ \ \ \ \ \ 
\ \ \ \ \ \ \ \ \ \ \ \ \ \ \ \ \ \ \ \ \ \ \ \ \ \ \ \ \ \ 
\ \ \ \ \ \ \ \ \ \ \ \ \ \ \ \ \ \ \ \ \ \ \ \ \ \ \ \ \ \ \cr
e_i \simeq l_2-\mu^{-1} \sqrt{ -2\ln(m^{exp}_{\tau^\pm} / \rho) };
e_j \simeq l_c+\mu^{-1} \sqrt{ -2\ln(m^{exp}_{\mu^\pm} / \rho) }; 
e_k \simeq l_a-\mu^{-1} \sqrt{ -2\ln(m^{exp}_{e^\pm} / \rho) }, \cr
B_{1,2}: 
\ \ \ \ \ \ \ \ \ \ \ \ \ \ \ \ \ \ \ \ \ \ \ \ \ \ \ 
\ \ \ \ \ \ \ \ \ \ \ \ \ \ \ \ \ \ \ \ \ \ \ \ \ \ \ \ \ \ 
\ \ \ \ \ \ \ \ \ \ \ \ \ \ \ \ \ \ \ \ \ \ \ \ \ \ \ \ \ \
\ \ \ \ \ \ \ \ \ \ \ \ \ \ \ \ \ \ \ \ \ \ \ \ \ \ \ \ \ \ \cr
e_i \simeq l_c-\mu^{-1} \sqrt{ -2\ln(m^{exp}_{e^\pm} / \rho) }; 
e_j \simeq l_a+\mu^{-1} \sqrt{ -2\ln(m^{exp}_{\tau^\pm} / \rho) }; 
e_k \simeq l_2+\mu^{-1} \sqrt{ -2\ln(m^{exp}_{\mu^\pm} / \rho) }, \cr
C_{1,2}: 
\ \ \ \ \ \ \ \ \ \ \ \ \ \ \ \ \ \ \ \ \ \ \ \ \ \ \ 
\ \ \ \ \ \ \ \ \ \ \ \ \ \ \ \ \ \ \ \ \ \ \ \ \ \ \ \ \ \ 
\ \ \ \ \ \ \ \ \ \ \ \ \ \ \ \ \ \ \ \ \ \ \ \ \ \ \ \ \ \
\ \ \ \ \ \ \ \ \ \ \ \ \ \ \ \ \ \ \ \ \ \ \ \ \ \ \ \ \ \ \cr
e_i \simeq l_c-\mu^{-1} \sqrt{ -2\ln(m^{exp}_{e^\pm} / \rho) }; 
e_j \simeq l_a+\mu^{-1} \sqrt{ -2\ln(m^{exp}_{\tau^\pm} / \rho) }; 
e_k \simeq l_2-\mu^{-1} \sqrt{ -2\ln(m^{exp}_{\mu^\pm} / \rho) }, \cr
D_{1,2}: 
\ \ \ \ \ \ \ \ \ \ \ \ \ \ \ \ \ \ \ \ \ \ \ \ \ \ \ 
\ \ \ \ \ \ \ \ \ \ \ \ \ \ \ \ \ \ \ \ \ \ \ \ \ \ \ \ \ \ 
\ \ \ \ \ \ \ \ \ \ \ \ \ \ \ \ \ \ \ \ \ \ \ \ \ \ \ \ \ \
\ \ \ \ \ \ \ \ \ \ \ \ \ \ \ \ \ \ \ \ \ \ \ \ \ \ \ \ \ \ \cr
e_i \simeq l_c-\mu^{-1} \sqrt{ -2\ln(m^{exp}_{\tau^\pm} / \rho) }; 
e_j \simeq l_a+\mu^{-1} \sqrt{ -2\ln(m^{exp}_{e^\pm} / \rho) }; 
e_k \simeq l_2+\mu^{-1} \sqrt{ -2\ln(m^{exp}_{\mu^\pm} / \rho) }, \cr
E_{1,2}: 
\ \ \ \ \ \ \ \ \ \ \ \ \ \ \ \ \ \ \ \ \ \ \ \ \ \ \ 
\ \ \ \ \ \ \ \ \ \ \ \ \ \ \ \ \ \ \ \ \ \ \ \ \ \ \ \ \ \ 
\ \ \ \ \ \ \ \ \ \ \ \ \ \ \ \ \ \ \ \ \ \ \ \ \ \ \ \ \ \
\ \ \ \ \ \ \ \ \ \ \ \ \ \ \ \ \ \ \ \ \ \ \ \ \ \ \ \ \ \ \cr
e_i \simeq l_c-\mu^{-1} \sqrt{ -2\ln(m^{exp}_{\tau^\pm} / \rho) }; 
e_j \simeq l_a+\mu^{-1} \sqrt{ -2\ln(m^{exp}_{e^\pm} / \rho) }; 
e_k \simeq l_2-\mu^{-1} \sqrt{ -2\ln(m^{exp}_{\mu^\pm} / \rho) },
\cr
\label{eq:10regEL}
\end{eqnarray} 
with $e_i=0$.

Secondly, in the regions $A_1,B_1,C_1,D_1,E_1$, the parameters are 
such that ($\{m,n,p\}$ are distributed among $[1,2,3]$, and, $b=1,3$: 
see Eq.(\ref{eq:rangeli})),
\begin{equation}
N_m \approx {3l_b-l_2 \over 2}, \
N_n \approx N_p \approx {l_b+l_2 \over 2},
\label{eq:5regLN}
\end{equation}
whereas for $A_2,B_2,C_2,D_2,E_2$, one has (with the same indice notation),
\begin{equation}
N_m \approx {3l_b-l_2 \over 2}, \
N_n \approx {l_b+l_2 \over 2}, \
N_p \approx {3l_2-l_b \over 2}. 
\label{eq:5regLNBIS}
\end{equation}
Note that both of these characteristic types of displacement configurations 
(\ref{eq:5regLN}) and (\ref{eq:5regLNBIS}) give rise to the pattern (same
indice convention):
\begin{equation}
\vert N_m-l_b \vert \approx 
\vert N_n-l_b \vert \approx
\vert N_n-l_2 \vert \approx
\vert N_p-l_2 \vert \approx
\vert {l_b-l_2 \over 2} \vert.
\label{eq:pattern}
\end{equation}

Thirdly, the field positions of lepton doublets satisfy to ($b=1,3$),
\begin{equation}
0.5 \lesssim \vert l_1-l_3 \vert \lesssim 1
\ \mbox{in} \ A_i,B_i,C_i,D_i,E_i \ [i=1,2], 
\label{eq:rangeli}
\end{equation}
\begin{eqnarray}
6.5 \lesssim \vert l_2-l_b \vert \lesssim 13.5 
\ \mbox{in} \ A_1,B_1,C_1,D_1,E_1, \cr
6.5 \lesssim \vert l_2-l_b \vert \lesssim 11.7
\ \mbox{in} \ A_2,B_2,C_2,D_2,E_2.
\label{eq:rangeliBIS}
\end{eqnarray}
In Eq.(\ref{eq:rangeliBIS}), the lower limit ($\sim 6.5$) is related to  
the experimental values for charged lepton masses (see examples of
configurations given in Section \ref{subsec:CLmasses3F}). In contrast, the 
existence of an upper limit ($\sim 13.5$ for configuration (\ref{eq:5regLN}) 
of parameters $l_i,N_j$ and $\sim 11.7$ for (\ref{eq:5regLNBIS})) in 
Eq.(\ref{eq:rangeliBIS}) is caused by the experimental constraints on 
mixing angles and neutrino mass squared differences 
(Eq.(\ref{eq:theta13exp}), Eq.(\ref{eq:Dmc23exp}) and 
Eq.(\ref{eq:Dmc12exp})).

In fact, each of the domains $A_i,B_i,C_i,D_i,E_i \ [i=1,2]$ in
$\{e,l,N,M_R\}$, reproducing all the present experimental data
on leptons, has a multiple nature. In the sense that there is some 
freedom in the distribution of $e$ indices $\{i,j,k\}$, $l$ indices 
$\{a,c\}$ (in Eq.(\ref{eq:10regEL})) and $N$ indices $\{m,n,p\}$ (in 
Eq.(\ref{eq:5regLN}) and Eq.(\ref{eq:5regLNBIS})). The fact, that the 
choice of distribution for $e$ indices $\{i,j,k\}$ is arbitrary, is due 
to the existence of an exact and formal symmetry, as explained in 
Section \ref{subsec:CLmasses3F}. In contrast, the invariance of solutions 
under any redistribution of the $N$ indices $\{m,n,p\}$, as well as the 
possibility to exchange $l_1$ and $l_3$ in regions 
$A_i,B_i,C_i,D_i,E_i \ [i=1,2]$, originate from accidental symmetries 
of the lepton mass matrices. In the sense that those non-trivial 
symmetries are caused by the specific structure of final solutions in 
$\{e,l,N,M_R\}$, and are thus correlated to the reproduced experimental 
results.

Finally, we mention that other similar solutions in $\{e_i,l_j,N_k,M_R\}$, 
fitting all the experimental results on the entire leptonic sector, can be 
obtained (see Eq.(\ref{eq:10regEL})) from the domains $A_{1,2}$ via the mass 
permutation $m^{exp}_{\mu^\pm} \leftrightarrow m^{exp}_{\tau^\pm}$,
from $B_{1,2}$ via $m^{exp}_{\mu^\pm} \leftrightarrow m^{exp}_{e^\pm}$ or
$m^{exp}_{\mu^\pm} \leftrightarrow m^{exp}_{\tau^\pm}$, from $C_{1,2}$ via 
$m^{exp}_{\mu^\pm} \leftrightarrow m^{exp}_{\tau^\pm}$, from $D_{1,2}$ via
$m^{exp}_{\mu^\pm} \leftrightarrow m^{exp}_{\tau^\pm}$ and from $E_{1,2}$ 
via $m^{exp}_{\mu^\pm} \leftrightarrow m^{exp}_{e^\pm}$ or
$m^{exp}_{\mu^\pm} \leftrightarrow m^{exp}_{\tau^\pm}$. Nevertheless, 
from any point of view, these new field position configurations 
do not differ significantly from the ones from which they are derived.

\end{document}